\documentclass[onecolumn]{svjour3}

\journalname{PLoS Computational Biology}

\usepackage{graphicx}
\usepackage{amsmath}
\usepackage{amssymb}
\usepackage{psfrag}
\usepackage{color}
\usepackage[nice]{nicefrac}
\usepackage{mathtools}
\usepackage[linesnumbered,ruled,lined]{algorithm2e}
\usepackage[right]{lineno}

\usepackage[
    backend=biber,
    style=nature,
    sorting=nyt
  ]{biblatex}
\RequirePackage{fix-cm}
\smartqed

\newcommand{\heading}[1]{\medskip\noindent\emph{#1}}
\newcommand{\ddt}[1]{\frac{{\rm d}{#1}}{{\rm d}t}}
\newcommand{\cR}{\mathcal{R}}
\newcommand{\E}{\mathbb E}
\newcommand{\prob}{\mathbf{P}}
\newcommand{\expect}{\operatorname{E}\expectarg}
\DeclarePairedDelimiterX{\expectarg}[1]{[}{]}{%
  \ifnum\currentgrouptype=16 \else\begingroup\fi
  \activatebar#1
  \ifnum\currentgrouptype=16 \else\endgroup\fi
}
\newcommand{\innermid}{\nonscript\;\delimsize\vert\nonscript\;}
\newcommand{\activatebar}{%
  \begingroup\lccode`\~=`\|
  \lowercase{\endgroup\let~}\innermid 
  \mathcode`|=\string"8000
}

\begin{document}

\title{Model Reduction Captures Stochastic Gamma Oscillations on Low-Dimensional Manifolds}% Low-Dimensional Manifolds Capture Gamma Oscillations with Model Reduction Methods OR Markovian Model Reduction Captures Stochastic Gamma Oscillations on Low-Dimensional Manifolds
\author{Yuhang Cai$^{1,*}$ \and Tianyi Wu$^{2,3,*}$ \and Louis Tao$^{3,4,\dagger}$ \and Zhuo-Cheng Xiao$^{5,\dagger}$}

\institute{  $^1$ Department of Statistics, University of Chicago, IL, USA 60637; \\
             $^2$ School of Mathematical Sciences, Peking University, Beijing 100871, China; \\
             $^3$ Center for Bioinformatics, National Laboratory of Protein Engineering and Plant Genetic Engineering, School of Life Sciences, Peking University, Beijing, 100871, China; \\
             $^4$ Center for Quantitative Biology, Peking University, Beijing, 100871, China; \\
             $^5$ Courant Institute of Mathematical Sciences, New York University, NY, USA 10003. \\
             $^*$       These two authors contributed equally to this paper and are listed in alphabetical order. \\
             $^\dagger$ Corresponding authors. 
              L.~Tao \email{taolt@mail.cbi.pku.edu.cn}; Z.-C.~Xiao \email{xiao.zc@nyu.edu}
}

\date{Received: date / Accepted: date }

\maketitle

%\large
%%%%%%%%%%%%%%%%%%%%%%%%%%%%%%%%%%%%%%%%%%%%%%%%%%%%%%%%%%%%
\begin{abstract}
Gamma frequency oscillations (25-140 Hz), observed in the neural activities within many brain regions, have long been regarded as a physiological basis underlying many brain functions, such as memory and attention. Among numerous theoretical and computational modeling studies, gamma oscillations have been found in biologically realistic spiking network models of the primary visual cortex. However, due to its high dimensionality and strong nonlinearity, it is generally difficult to perform detailed theoretical analysis of the emergent gamma dynamics. Here we propose a suite of Markovian model reduction methods with varying levels of complexity and apply it to spiking network models exhibiting heterogeneous dynamical regimes, ranging from nearly homogeneous firing to strong synchrony in the gamma band. The reduced models not only successfully reproduce gamma oscillations in the full model, but also exhibit the same dynamical features as we vary parameters.  Most remarkably, the invariant measure of the coarse-grained Markov process reveals a two-dimensional surface in state space upon which the gamma dynamics mainly resides.  Our results suggest that the statistical features of gamma oscillations strongly depend on the subthreshold neuronal distributions.  Because of the generality of the Markovian assumptions, our dimensional reduction methods offer a powerful toolbox for theoretical examinations of other complex cortical spatio-temporal behaviors observed in both neurophysiological experiments and numerical simulations. %\tnote{@Zhuocheng here describe one interesting finding before mentioning invariant measure}  \tnote{@Zhuocheng maybe one more sentence about this result before going to the concluding sentence.} 

\end{abstract}





\section{Author Summary}
Emergent oscillations in the brain may be the neural basis for many brain functions, such as memory and attention. In particular, gamma band oscillations have been found concomitant with improvements in sensory and cognitive tasks (and its disruption have been associated with various brain disorders). A major goal of theoretical and computational neuroscience is to understand how complex dynamical neural activities can emerge from the interplay of cellular and network elements. Here we present a sequence of reduction methods to examine the development of gamma dynamics in an idealized Integrate-and-Fire network model. Through coarse-graining, we find that the appearance of gamma band oscillations can be represented on a suitable, low-dimensional manifold. The reduced systems can recapitulate many of the statistics of the full model, including the transition from homogeneous firing to gamma synchrony. Furthermore, our coarse-graining reveals the importance of subthreshold neuronal distributions that may account for the emergence of gamma oscillations in the brain.

%%\linenumbers

%%%%%%%%%%%%%%%%%%%%%%%%%%%%%%%%%%%%%%%%%%%%%%%%%%%%%%%%%%%%
\section{Introduction\label{sec_intro}}
%%% Introduction
Modern experimental techniques have revealed a vast diversity of coherent spatiotemporal activity patterns in the brain, reflecting the many possible interactions between excitation and inhibition, between cellular and synaptic time-scales, and between local and long-range circuits. Prominent amongst these patterns are the rich repertoire of neuronal oscillations that can be stimulus driven or internally generated and are likely to be responsible for sensory perception and cognitive tasks \cite{Fries2009,TallonBaudry2009}. In particular, gamma band oscillations (25-140 Hz), observed in multi-unit activity (MUA) and local field potential (LFP) measurements \cite{RayMaunsell2015}, have been found in many brain regions (visual cortex \cite{GrayEtAl1989,azouz2000dynamic,LogothetisEtAl2001,HenrieShapley2005}, auditory cortex \cite{BroschEtAl2002}, somatosensory cortex \cite{bauer2006tactile}, parietal cortex \cite{PesaranEtAl2002,BuschmanMiller2007,MedendorpEtAl2007}, frontal cortex \cite{BuschmanMiller2007,GregorgiouEtAl2009,SiegelEtAl2009,
SohalEtAl2009,CanoltyEtAl2010,SigurdssonEtAl2010,van2010learning}, hippocampus \cite{bragin1995gamma,CsicsvariEtAl2003,ColginEtAl2009,colgin2016rhythms}, amygdala \cite{PopescuEtAl2009} and striatum \cite{vanderMeerRedish2009}). Much experimental evidence correlates gamma oscillations to behavior and enhanced sensory or cognitive performances. For instance, gamma dynamics has been shown to sharpen orientation tuning in V1 and speed and direction tuning in MT \cite{azouz2000dynamic,AzouzGray2003,FrienEtAl2000,
WomelsdorfEtAl2012,LiuNewsome2006}. During cognitive tasks, the increases in gamma power in the visual pathway have been shown to correlate with attention \cite{FriesEtAl2001,FriesEtAl2008}. Experiments have implicated gamma oscillations during learning \cite{bauer2007gamma} and memory \cite{PesaranEtAl2002}. Numerical studies have demonstrated that coherent gamma oscillations between neuronal populations can provide temporal windows during which information transfer can be enhanced \cite{WomelsdorfEtAl2007}. The disruption of gamma frequency synchronization is also concomitant with multiple brain disorders \cite{Basar2013,bressler2003context,mcnally2016gamma,krystal2017impaired,mably2018gamma}.

Many large-scale network simulations \cite{TraubEtAl2005,chariker2015emergent} and firing rate models \cite{BrunelHakim1999,KeeleyEtAl2019} have been used to capture the wide range of the experimentally observed gamma band activity. The main mechanism underlying the emergent gamma dynamics appears to be the strong coupling between network populations, either via synchronizing inhibition \cite{whittington2000inhibition} or via competition between excitation and inhibition \cite{whittington2000inhibition,chariker2015emergent}. However, a theoretical account of how the collective behavior emerges from the detailed neuronal properties, local network properties and cortical architecture remains incomplete.

Recently, Young and collaborators have examined the dynamical properties of gamma oscillations in a large-scale neuronal network model of monkey V1 \cite{chariker2018rhythm}. To further theoretical understanding, in \cite{li2019well}, Li, Chariker and Young introduced a relatively tractable stochastic model of interacting neuronal populations designed to capture the essential network features underlying gamma dynamics. Through numerical simulations and analysis of three dynamical regimes (``homogeneous", ``regular" and ``synchronized"), they identified how conductance properties (essentially, how long after each spike the synaptic interactions are fully felt) can regulate the emergence of gamma frequency synchronization.

Here, we present a sequence of model reductions of the spiking network models based on Li, Chariker and Young \cite{li2019well}. We first present in detail our methods on a small, homogeneous network of 100 neurons (75 excitatory and 25 inhibitory), exhibiting gamma frequency oscillations.  Inspired by \cite{li2019well,li2019stochastic}, to achieve dimensional reduction, we assume that the spiking activities during gamma oscillations and their temporally organization are mainly governed by one simple variable, namely, which sub-threshold neurons are only a few postsynaptic spikes from firing.  Thus, in terms of dynamical dimensions, the number of network states is drasticaly reduced (although still too large for any meaningful analytical work, since $2^n$ is astronomical even for $n = 100$).  Therefore, we further coarse-grain by keeping count of the numbers of neurons that are only a few postsynaptic spikes from threshold. The number of effective states is then reduced to the order of millions.  By restricting the dynamics onto this dimensionally-reduced state space, it is now possible to make use of the classical tools of stochastic models to analyze the emergence and statistical properties of gamma frequency synchronization. The reduced models not only successfully capture the key features of gamma oscillations, but strikingly, they also reveal a simple, low-dimensional manifold structure of the emergent dynamics.
%\medskip

The outline of this paper is as follows.  In Section~\ref{sect_Rslt}, we present our model reductions, i.e., the sequence of models going from the full model, to the two-state, reduced network model and to its coarse-grained simplification.  Especially, we show the low-dimensional manifold representing the gamma dynamics.  We discuss the connection of our work to previous studies and how it may benefit future research of gamma dynamics in Section~\ref{sec_discs}.  Computational methods and technical details are provided in the Methods section.

%%%%%%%%%%%%%%%%%%%%%%%%%%%%%%%%%%%%%%%%%%%%%%%%%%%%%%%%%%% Fig1
\begin{figure}
  \begin{center}
    %% \captionsetup{type=figure} 
    \includegraphics[width=0.85\textwidth]{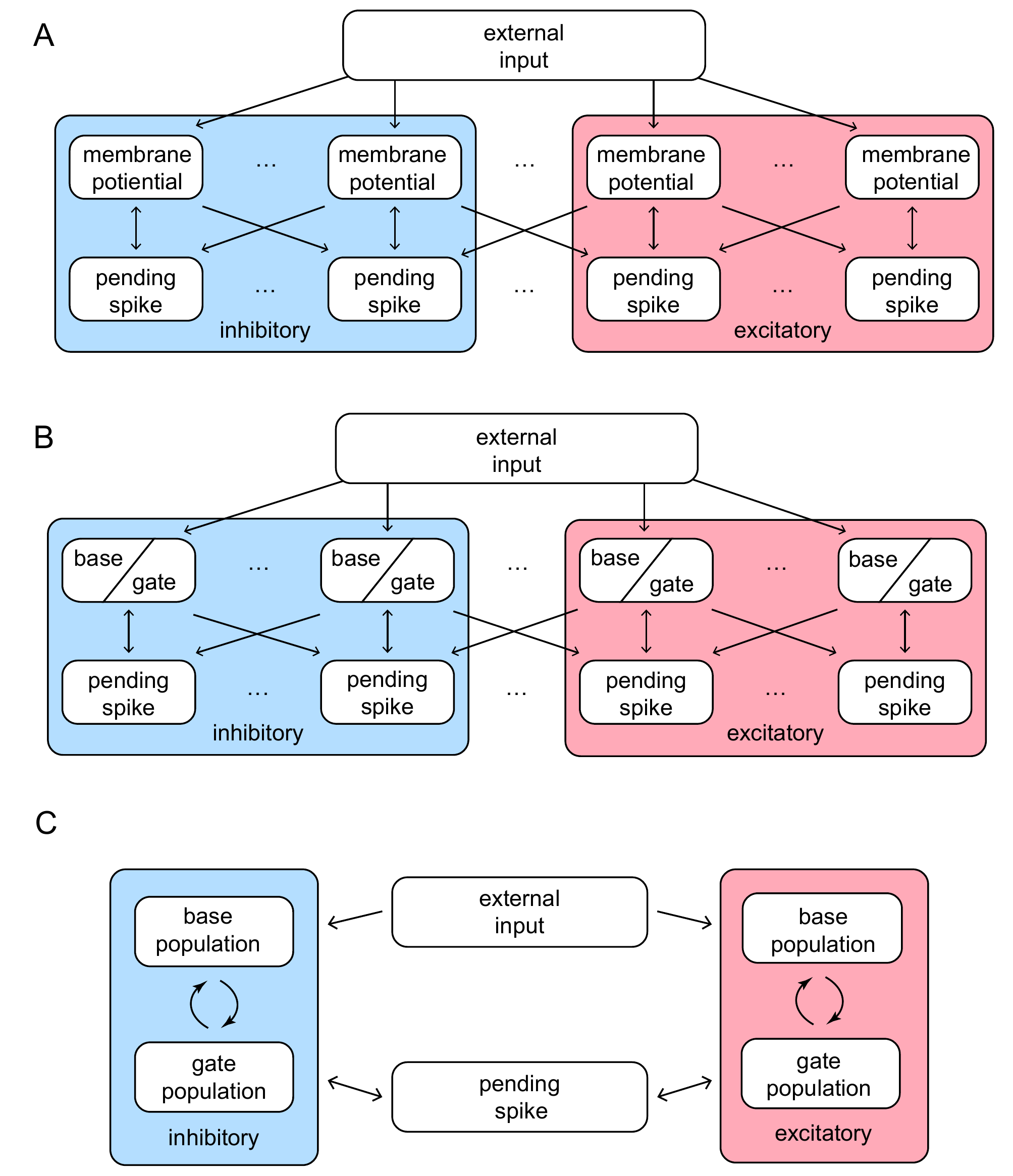}
    \caption{Structures and important features of 3 models.  \textbf{A.} The structure of the full Markovian integrate-and-fire network.  \textbf{B.} The structure of the reduced network model, where the membrane potentials only take values in $\{\mbox{Base, Gate}\}$. \textbf{C.} The structure of coarse grained model. In this model the pending $E$-kick pools for every neuron are merged in to one for the whole network, and so are the pending $I$-kick pools. } % \textbf{D.} The spiking raster plot and several associated quantities for synchronous regime in full model, including pending spikes $(H_{EE}, H_{EI}, H_{EI}, H_{II})$ and the portion of gate neurons ($N_{GE}/N_{E}, N_{GI}/N_I$). These quantities are employed to investigate the consistency of three different level models for each regime. \textbf{E.} Firing Cluster dissection plot for the same simulation shown in \textbf{D}.
    \label{Fig1: Model} % use \ref{fig1} to reference to this figure
  \end{center}
\end{figure}
%%%%%%%%%%%%%%%%%%%%%%%%%%%%%%%%%%%%%%%%%%%%%%%%%%%%%%%%%%% Fig1

%%%%%%%%%%%%%%%%%%%%%%%%%%%%%%%%%%%%%%%%%%%%%%%%%%%%%%%%%%%%
\section{Results\label{sect_Rslt}}
In spiking neuronal networks, gamma frequency oscillations appear as temporally repeating, stochastic spiking clusters in the firing patterns.  Different mechanisms have been proposed to explain this phenomena.  As a minimal mechanism, interneuron network gamma (ING) proposes that the gamma oscillations can be produced by the fast interactimodelingons between inhibitory ($I$) neurons alone \cite{whittington2000inhibition}. When inhibitory neurons have intrinsic firing rates higher than the gamma band, they may exhibit gamma firing frequency when mutually inhibited. ING does not require the existence of excitatory ($E$) neurons but studies showed that it loses its coherence in systems where neurons are heterogeneously driven \cite{wang1996gamma}. Another class of theories view the repeating collective spiking clusters as the outcome of competition between $E$ neurons and $I$ neurons, such as the pyramidal-interneuron network gamma (PING, \cite{whittington2000inhibition,borgers2003synchronization}) and recurrent excitation-inhibition (REI, \cite{chariker2018rhythm}). % \tnote{Say a little about how ING works} \tnote{rewrite first sentence} \tnote{can we be less negative here?}

Though sharing many common qualitative features, PING and REI provide substantially different explanations to the formation of gamma oscillations. Most remarkably, the collective spiking clusters in PING are usually whole-population spikes as a result of steady external input.  The $E$-population spikes induce $I$-population spiking activities that are offset in time, and strong enough to suppress the entire network.  A new $E$-population spiking event occurs after the inhibition wears off, leading to a series of nearly periodic, whole-population oscillatory activity.

On the other hand, the REI mechanism depicts a highly stochastic network dynamics: Driven by noisy stimulus, a few $E$-neurons crosses the threshold, and the subsequent recurrent excitation recruit more spikes from other excitatory neurons, leading to rapidly rising spike clusters. But this can not go forever since a few of the inhibitory neurons are excited at the same time, pushing the whole population to a less excitable condition. The next collective spiking event then emerges from the victory of excitation during its competition with inhibition.  Therefore, the important features of the transient spike clusters are highly temporally variable, and the gamma frequency band of the oscillation is mainly a statistical feature of the emergent complex temporal firing patterns.  %\tnote{@Zhuocheng This section focused too much on MFEs which not every one agrees to be the ``cause" of gamma. Let us think about how to re-write this paragraph.} % Moreover, \cite{chariker2018rhythm} reveals that the feedback interactions within the network is the primary source of the input to each neuron, whereas the feedforward external stimulus can only contributes $<10\%$. 

Our primary goal is to provide a better understanding of the emergence of gamma oscillations from the interaction between neurons in this nonlinear, stochastic, and high-dimensional context. In general, the biggest difficulty of analyzing spiking network dynamics is its dimensionality. Consider a network of $N$ neurons, the number of possible states grows exponentially as $N\to\infty$, no matter if we use a single-neuron model as complex as the Hodgkin-Huxley \cite{hodgkin1952quantitative} or as simple as binary-state \cite{cowan1991stochastic}.  Previously, many attempts have been made for model reduction by capturing collective network dynamics in specific dynamical regimes.  Successful examples include kinetic theories \cite{CaiKinetic2006} for mean-field firing patterns, statistical field theories \cite{buice2013beyond} for higher-order correlations, Wilson-Cowan neural field model \cite{wilson1972excitatory,wilson1973mathematical} for spatial-temporal patterns, to name a few. In this study, our reduced models are developed with similar motivations to capture gamma oscillations. Generally, models incorporating many biologically realistic details can be very complicated; The reduced models are much easier to analyze, but some of the neglected information can lead to biases in many aspects. Here, we aim for a balance between realism and abstraction. Specifically, we present a sequence of reduced models, between which the connections are well defined (and can be mathematically analyzed in future studies). Importantly, we find that even the coarsest model preserves important features and statistics of the gamma oscillations found in the full model. 

Our reduced models are based on the integrate-and-fire (IF) model, which is widely employed in previous models of spiking networks (see Methods).  In a recent study, gamma oscillations have been found emerging from the simulation of a large-scale IF network model of layer 4C$\alpha$ of V1 \cite{chariker2018rhythm}.  Later theoretical studies suggest that gamma oscillations in different spatial regions decorrelate quickly on the scale of a couple of hypercolumns, echoing experimental observations that gamma oscillation is very local in cortex \cite{goddard2012gamma,lee2003synchronous,menon1996spatio}.  Therefore, we first focus on a small homogeneous network with 75 excitatory neurons ($E$) and 25 inhibitory neurons ($I$) as an analogy of an orientation-preference domain of a hypercolumn in V1.  

%%%%%%%%%%%%%%%%%%%%%%%%%%%%%%%%%%%%%%%%%%%%%%%%%%%%%%%%%%% Fig2
\begin{figure}
  \begin{center}
    %% \captionsetup{type=figure} 
    \includegraphics[width=\textwidth]{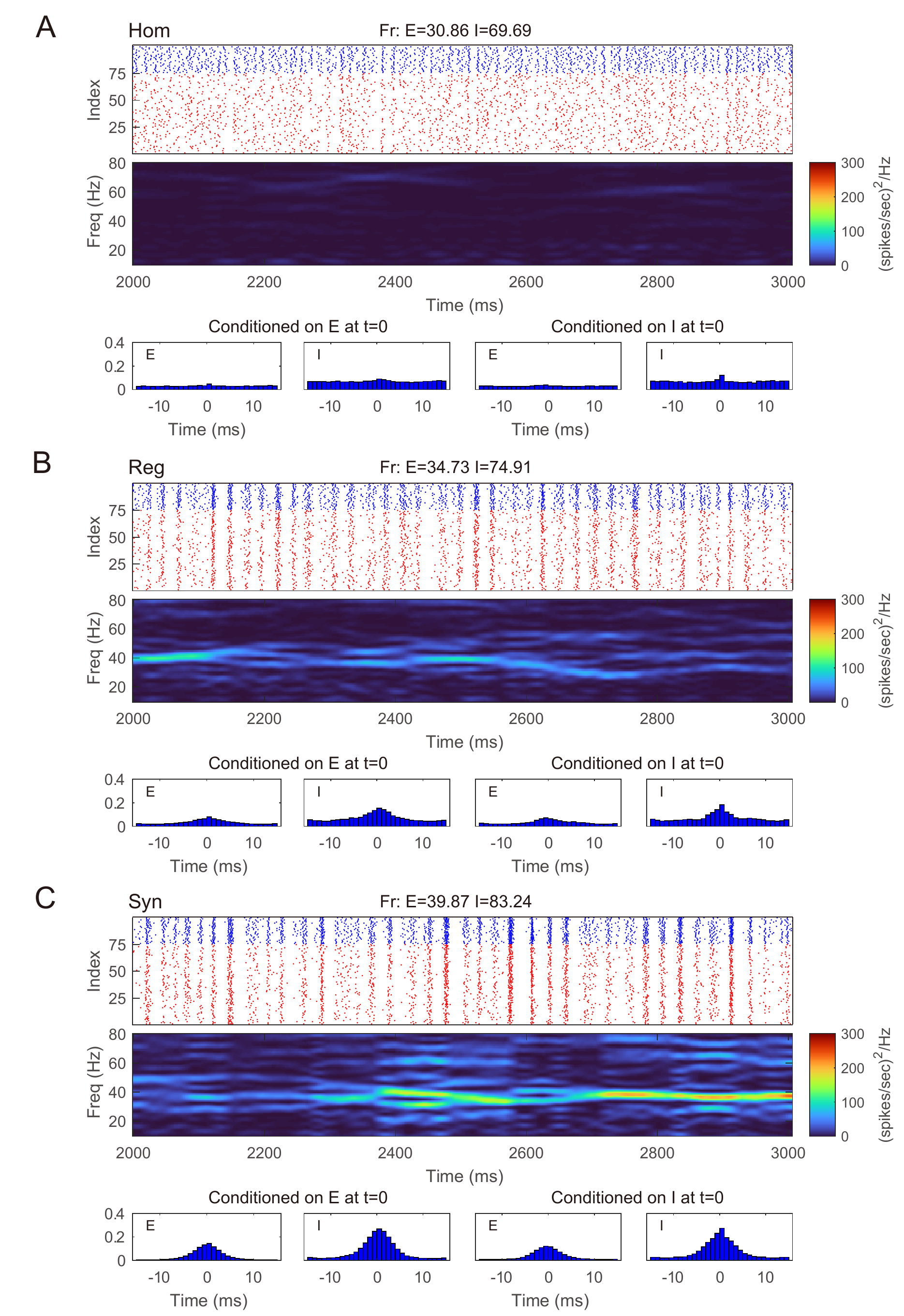}
    \caption{Gamma oscillation exhibited by the MIF network in three different regimes.  \textbf{A.} Homogeneous regime (``Hom"). Top: Raster plots of all 75 $E$-neurons (blue) and $I$-neurons (red) with firing rates noted in title. Middle: The spectrogram of the firing pattern exhibiting low density in the gamma band (30-80 Hz). Bottom: The spiking correlation diagrams for $\{E,I\}$-spikes.  \textbf{B and C.} Same as \textbf{A}. The regular (``Reg") and synchronized (``Syn") regimes exhibit much more synchronous firing patterns and stronger gamma-band activity.} % \textbf{D.} The spiking raster plot and several associated quantities for synchronous regime in full model, including pending spikes $(H_{EE}, H_{EI}, H_{EI}, H_{II})$ and the portion of gate neurons ($N_{GE}/N_{E}, N_{GI}/N_I$). These quantities are employed to investigate the consistency of three different level models for each regime. \textbf{E.} Firing Cluster dissection plot for the same simulation shown in \textbf{D}.
    \label{Fig2: MIF} % use \ref{fig1} to reference to this figure
  \end{center}
\end{figure}
%%%%%%%%%%%%%%%%%%%%%%%%%%%%%%%%%%%%%%%%%%%%%%%%%%%%%%%%%%% Fig2

\subsection{Reduced Models Captures Statistics of Gamma Oscillations}
\subsubsection{A Markovian Intergrate-and-Fire network}
We start with the Markovian integrate-and-fire model (MIF) first proposed in \cite{li2019well}, and hereafter referred to as the ``full model."  As an analogy of the conventional IF model (see Methods), the MIF brings us two additional conveniences: First, the discretized states of Markovian dynamics make theoretical analysis easier as the probability flow from one state to another is now straightforward; Second, the Markov properties of the MIF enable the computation of the invariant measure of gamma oscillations directly from the probability transition matrix.

Our MIF model network is composed of 100 interacting neurons (75 $E$-neurons and 25 $I$-neurons) driven by external Poissonian stimulus (Fig.~\ref{Fig1: Model}A). The state of neuron $i$ is described by three variables $(v_i,H^E_i,H^I_i)$, where $v_i$ represents individual membrane potentials and $H^{\{E,I\}}_i$ are analogies of the $E$ and $I$ conductances (see below). We use the size of external kicks to discretize the membrane potential, and let $v_i$ range from $V^I = -66$, the reversal potential of inhibitory synapses, to $V^{th} = 100$, the spiking threshold. Immediately after reaching $V^{th}$, $v_i$ enters the refractory state $\cR$, and, at the same time, sends a spike to its postsynaptic targets. After an exponentially distributed waiting-time $\tau^\cR$, $v_i$ resets to the rest potential $V^r = 0$. The `integrate' part of MIF is separated into two components, external and recurrent. Each external kick increases $v_i$ by 1. To model the effects of recurrent network spikes, we use $H^{\{E,I\}}_i$ to denote the number of postsynaptic spikes (forming `pools' of pending-kicks) received by neuron $i$ that has not yet taken effect. When receiving an ${\{E,I\}}$-spike, the corresponding pending-kick pool, $H^{\{E,I\}}_i$, increases by 1. Each postsynaptic spike affects $v_i$ independently, and after an exponentially distributed waiting-time (i.e., the synaptic time-scale) $\tau^{\{E,I\}}$, increase or decrease $v_i$ (depending on $\{E,I\}$ of the presynaptic neuron). The specific increment/decrement depends on the synaptic strength and the state of $v_i$. The connections between neurons are homogeneous: Whether a spike released by neuron $i$ is received by neurons $j$ is determined by an independent coin flip, whose probability only depends on the type of neuron $i$ and $j$. We leave the details and choices of parameters to Methods. 

This 100-neuron MIF network exhibits gamma-band oscillations as demonstrated in \cite{li2019well} (Fig.~\ref{Fig2: MIF}).  By varying the synaptic time scales $\tau^{\{E,I\}}$, we examine three regimes with different degrees of synchrony: homogeneous (``Hom"), regular (``Reg"), and synchronized (``Syn"). (Specifically, we fix the expectation of the waiting time for $I$-kicks ($\tau^I$) and manipulate separately the expectation of waiting time for $E$-spikes on $E$ and $I$ neurons, $\tau^{EE}$ and $\tau^{IE}$; for full details, see Methods). In the raster plot of ``Hom'' regime, the MIF network produces a firing pattern in which spikes do not exhibit strong temporal correlations (Fig.~\ref{Fig2: MIF}A top). (This is also verified by the spike-time correlations conditioned on each $\{E,I\}$-spike; see Fig.~\ref{Fig2: MIF}A bottom). Meanwhile, no strong spectral density peak is seen in the spectrogram (Fig.~\ref{Fig2: MIF}A middle). In the ``Reg" regime, however, spikes begin to cluster in time as multiple firing events (hereafter, MFEs)  and exhibit stronger spike-time correlations \cite{RanganYoung2013a,RanganYoung2013b}. Namely, MFE is a temporally transient phenomenon lying between homogeneity and total synchrony, where a part (but not all) of the neuronal population fires during a relatively short time window, as widely reported in previous experimental and modelling studies \cite{beggs2003neuronal,churchland2010stimulus,plenz2011multi,shew2011information,yu2010membrane,yu2011higher}. Affected by MFEs, the mass of the spectral density is primarily located in the gamma band, especially around 40-60 Hz (Fig.~\ref{Fig2: MIF}B).  Many more and stronger synchronized firing patterns, higher spike-time correlations, and stronger gamma-band spectral peaks are observed in the ``Syn" regime.  These dynamics and statistics are consistent with the results of \cite{li2019well} and in IF neuronal network simulations \cite{ZhangNewhallEtAl2014,ZhangZhouEtAl2014,ZhangRangan2015}. Furthermore, the gamma-band spectrograms are comparable with experimental studies \cite{xing2012stochastic}. %\tnote{Describe MFEs here.}

We note that although our MIF network only consists of $N = 100$ neurons, the number of states in the Markov chain is $(168\cdot n_{H^E}\cdot n_{H^I})^{N}$, where $n_{\{H^E,H^I\}}$ are the largest possible sizes of the pending spike pools (see Methods for precise definition).  Therefore, it is computationally unrealistic to do any meaningful analytical work to understand the dynamics, especially how gamma oscillations can emerge from the probability flows between different states. Therefore, we regard this MIF network as the ``full model," and apply dimensional reduction methods for further analysis.

%%%%%%%%%%%%%%%%%%%%%%%%%%%%%%%%%%%%%%%%%%%%%%%%%%%%%%%%%%% Fig3
\begin{figure}
  \begin{center}
    %% \captionsetup{type=figure} 
    \includegraphics[width=\textwidth]{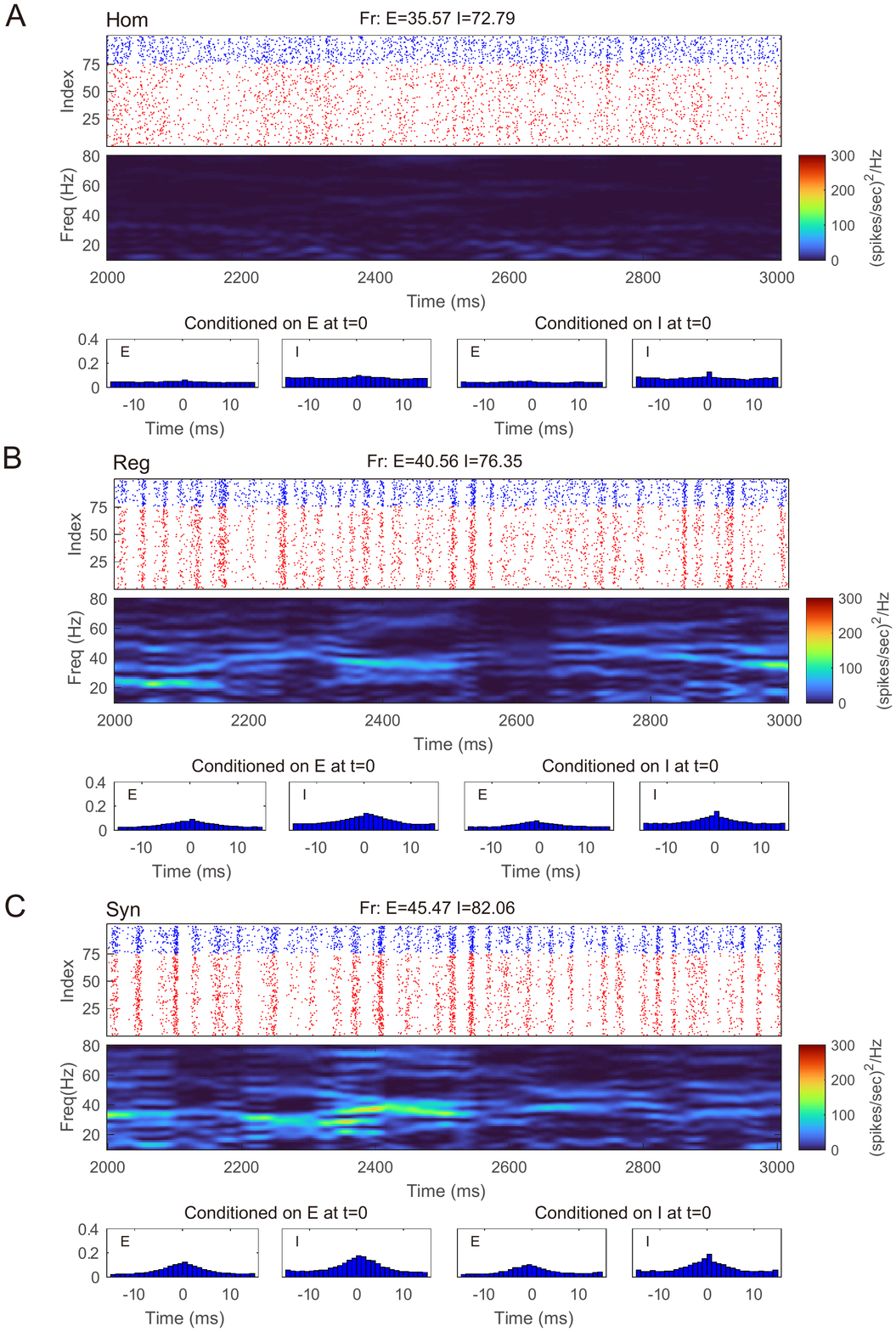}
    \caption{Gamma oscillation captured by the reduced network consisting of two-state neurons.  \textbf{A-C.} Same regimes and statistics investigated in Fig.~\ref{Fig2: MIF}.} 
    \label{Fig3: RN}
  \end{center}
\end{figure}
%%%%%%%%%%%%%%%%%%%%%%%%%%%%%%%%%%%%%%%%%%%%%%%%%%%%%%%%%%% Fig3

\subsubsection{A Reduced Network with Two-State Neurons}
First, we introduce a reduced network (RN) consisting of two-state neurons, i.e., RN model has the same setup as the MIF network except that the membrane potentials only have two states: \textit{base} and \textit{gate} (Fig.~\ref{Fig1: Model}B).  From the perspective of the full model, a neuron is deemed as a base or gate neuron by how far it is away from firing. Specifically, we set a cutoff $V^c$ below the threshold $V^{th}$, and neuron $i$ is regarded as a gate neuron if $v_i>V^c$, otherwise it is a base neuron (including $v_i\leq V^c$ and $v_i = \cR$). Neuron $i$ flips between the base and gate states when % \tnote{reduced MODEL or reduced NETWORK?}\xnote{@Louis, I am inclined to keep reduced network for now. The term "reduced models" includes both RN and CG in this paper.} 
\begin{enumerate}
    \item $v_i$ crosses the cutoff $V^c$, or
    \item Neuron $i$ fires and enters the refractory state $\cR$.
\end{enumerate}
However, the reduction to two-state neurons immediately raises a question: Without the consecutive discrete states between $[V^I,V^{th}]$, how can we represent the effect of (external, $E$/$I$-) kicks on individual neurons when $v_i$ only takes two possible states?  

Consider an $E$-neuron $i$ in the gate state, i.e., $v_i>V^c$ in the corresponding MIF network. Since we do not know the exact value \textit{a priori}, $v_i$ can take any value between $[V^c, V^{th}]$ with a probability determined by the empirical distribution of the whole $E$-population. When an $E$-kick takes effect and increases $v_i$ by a synaptic strength of $S_{EE}$, neuron $i$ fires and changes state to ``base" if and only if $v_i$ locates at the excitable region $(V^{th}-S_{EE}, V^{th}]$, otherwise it stays in the gate state (and $v_i\in[V^c,V^{th}-S_{EE})$). Therefore, a single $E$-kick has a probability
\begin{equation}
    P_{E}^{GE} = \prob\left(v_i\in(V^{th}-S_{EE}, V^{th}] | v_i\in(V^c, V^{th}]\right) = \frac{\prob(v_i\in(V^{th}-S_{EE}, V^{th}])}{\prob(v_i\in(V^c, V^{th}])},
\end{equation}
to excite a gate $E$-neuron, leading to its spike and transition to the base state. That is, $P_{E}^{GE}$ is the transition probability of neuron $i$ in the excitable region, conditioned on that neuron $i$ is a gate neuron. \textit{A priori}, we do not have the full distribution of neuronal states, therefore, in order to close the RN model, we use statistical learning methods by inferring $P_{E}^{GE}$ from a long-term simulation of the full MIF network. Likewise, we acquire all other transition possibilities induced by kicks (see Methods).

This reduction of classifying spiking neurons into two states is based on one simple assumption: The emergence of MFEs in the collective dynamics is mostly sensitive to one variable, i.e., the number of subthreshold neurons are only a few spikes from firing (gate neurons).  On the other hand, the distribution of neurons with lower membrane potentials (base neurons) is less immediately relevant since it is much less possible for them to generate spikes in the next few milliseconds.  Therefore, the initiation and maintenance of gamma oscillations are dominated by the probability flow between gate and base states.
%  First of all, one crucial insight provided by \cite{chariker2015emergent} is that the probability distribution of the membrane potentials stays close to the threshold $V^{th}$ and does not experience very sharp change throughout the repeating multiple firing events.  Therefore, it is reasonable to assume that the emergence of the spiking clusters in gamma oscillations is sensitive one simple variable

Even with such a drastic simplification, the RN model provides remarkably good approximation of dynamics produced by the MIF network (Fig.~\ref{Fig3: RN}), which is verified by raster plots, firing rates, and spectrum densities similar to those of the full model in Fig.~\ref{Fig2: MIF}.  On the other hand, we notice that the spike-time correlations in the ``Syn" regime are indeed slightly lower than the corresponding value in the full model.  This is not very surprising: When the spiking pattern is synchronized, the dynamics is more sensitive to the details of the probability distribution of neurons in the excitable region since one spike may trigger more spikes followed by other neurons.  One may, of course, consider using more information to describe the full distribution of the membrane potentials (perhaps by using three or more states instead of two).  Though a more detailed models may provide us with better numerical approximations, our primary goal of capturing the key features of gamma oscillation has been well served by the current version of the RN model with two-state neurons. % \tnote{Describe a little what happens; which features of the full model are captured by RN}

Though radically simplified, our RN model is still a Markov chain, with $(2\cdot n_{H^E}\cdot n_{H^I})^{N}$ states. However, the setup and success of the RN model provide important insights for further model reduction.

%%%%%%%%%%%%%%%%%%%%%%%%%%%%%%%%%%%%%%%%%%%%%%%%%%%%%%%%%%% Fig4
\begin{figure}
  \begin{center}
    %% \captionsetup{type=figure} 
    \includegraphics[width=\textwidth]{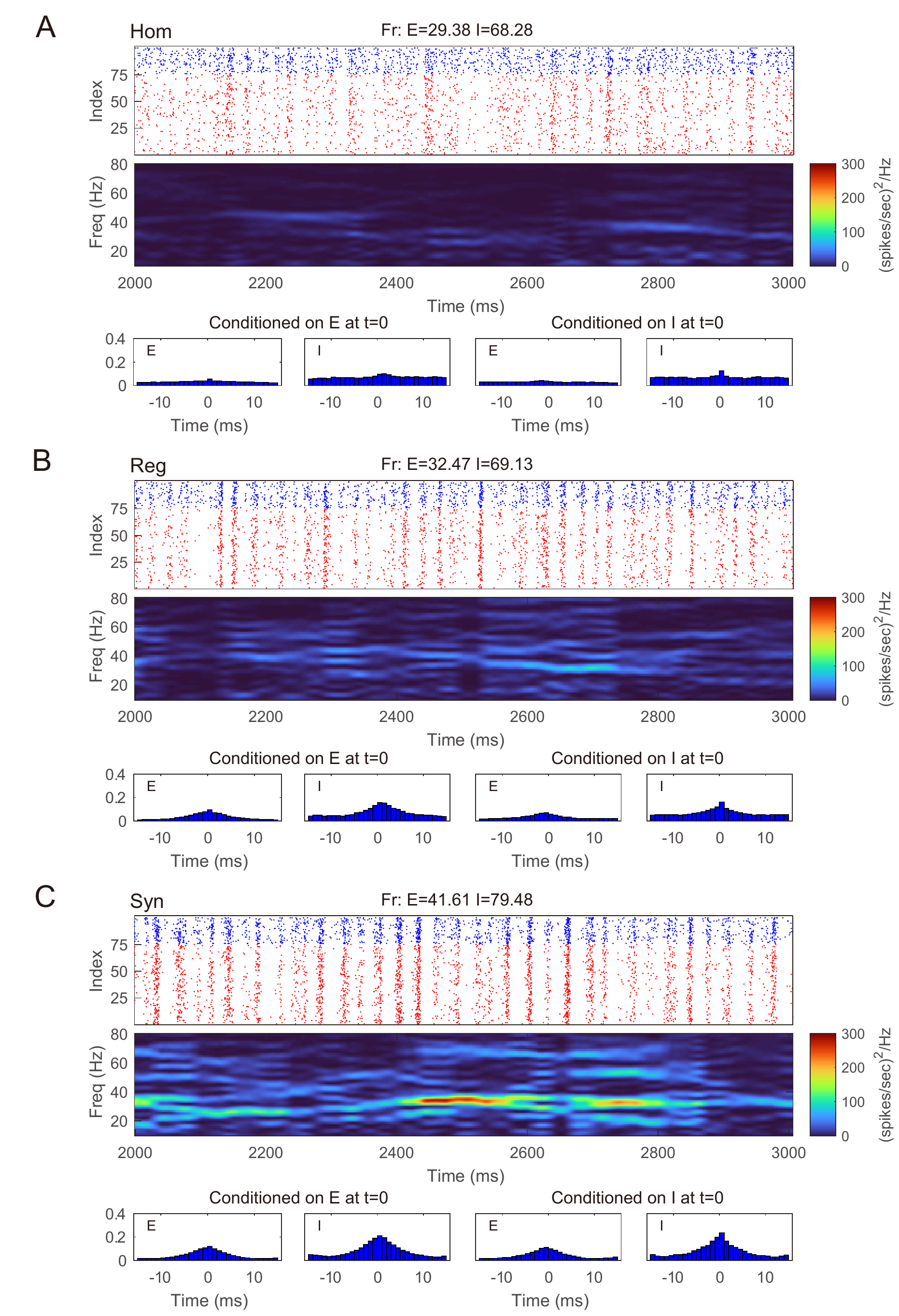}
    \caption{Gamma oscillation captured by the coarse-grained model.  \textbf{A-C.} Same regimes and statistics investigated in Fig.~\ref{Fig1: Model} and \ref{Fig2: MIF}. The fake raster plots are produced by randomly assigning spikes to each neuron.} 
    \label{Fig4: CG}
  \end{center}
\end{figure}
%%%%%%%%%%%%%%%%%%%%%%%%%%%%%%%%%%%%%%%%%%%%%%%%%%%%%%%%%%% Fig4

\subsubsection{A Coarse-Grained Approximation}
When we infer the transition probabilities between gate and base states, we are uncertain of the full distribution of $v_i$ in the network (besides the number of neurons in base and gate states).   Therefore, the success of the RN model suggests that we may think of the transition probabilities as functions of the number of $\{E,I\}$ neurons in each state: $N_{GE}$, $N_{GI}$, $N_{BE}$, $N_{BI}$.  Thus, the core idea of further simplifying by coarse-grained (CG) approximation is this: Instead of thinking about the state of each individual neuron, we study the state of population statistics. Below we summarize the setup of our CG model and leave the details to Methods.

Let us first consider the pending-kick pools of a single neuron.  Take the $I$-to-$E$ and $I$-to-$I$ projections as an example: Because of the homogeneity of the network, for an $I$-spike, each of its postsynaptic neuron receives it independently with a probability that depends only on the specific neuronal type $\{E,I\}$.  In addition, each spike takes effect independently, with the same waiting time distribution (exponential with mean $\tau^I$).  Therefore, this is equivalent to a collective $I$-kick pool of size $H^{I}$, representing the sum of all $I$-kick pools in the entire network.  In this pool, each pending-kick takes effect independently and is randomly distributed to a specific neuron with a probability depending only on its $E/I$ type.  With similar considerations, the $E$-kick pools are also merged into one at a size $H^E$. However, since $(\tau^{EE},\tau^{IE})$ are separately manipulated in the full model, we have to ignore the subtle differences between the consumption rates of pending $E$-kicks on $E$ and $I$ neurons. Specifically, we assume that a constant portion of $H^E$ are distributed to $E$ ($I$) neurons (i.e., $H^{EE} = a^{EE}\cdot H^E$. See Methods), which introduces a bias in our CG model.

Since each neuron is driven by the same, coarse-grained $\{E,I\}$-kick pools, due to the interchangeability of $I$ neurons, they are now only differentiated by their current state (base or gate). Similarly, the information of the state of every $I$ neuron is now represented by the Numbers $(N_{BI},N_{GI})$.  Since the total number of $I$ neurons is fixed, we only need $N_{GI}$. These considerations also apply to the $E$ neurons. Thus, our CG model becomes a Markov chain with four variables:
\begin{equation}
    (N_{GE},N_{GI},H^E,H^I).
\end{equation}
Kicks taking effect may change $N_{GE}$ and $N_{GI}$, and spikes released by neurons increase $H^E$ and $H^I$.  The CG model contains $N_E\cdot N_I\cdot N_{H^E}\cdot N_{H^I}$ states ($N_{H^E} = n_{H^E}\cdot N$, $N_{H^I} = n_{H^I}\cdot N$).  For each state, there are at most 12 possible transitions to other states (see Methods).  Therefore, the CG model becomes an $O(N^4)$ problem, allowing us to analyze the dynamics in detail.

The CG model is a natural simplification of the RN model, and it is reasonable to expect CG to capture the main features of gamma oscillations.  Indeed, we find that in all three regimes we considered, the behaviors of CG model are in well agreement with both the MIF and RN models (Fig.~\ref{Fig4: CG}).  Note that, since we do not track the dynamics of individual neurons in CG, the (fake) raster plots are generated by randomly assigning spikes to neurons.

%Qs:
%1. Why using small, homogenous network?  
%2. Why discretized-states Markovian?  
%3. Why binary states, but not three states? 
%4. Why coarse-grained? What has been omitted? 
%5. Regimes selected

\subsection{Gamma Dynamical Features in Reduced Models}
\begin{figure}
  \begin{center}
    %% \captionsetup{type=figure} 
    \includegraphics[width=\textwidth]{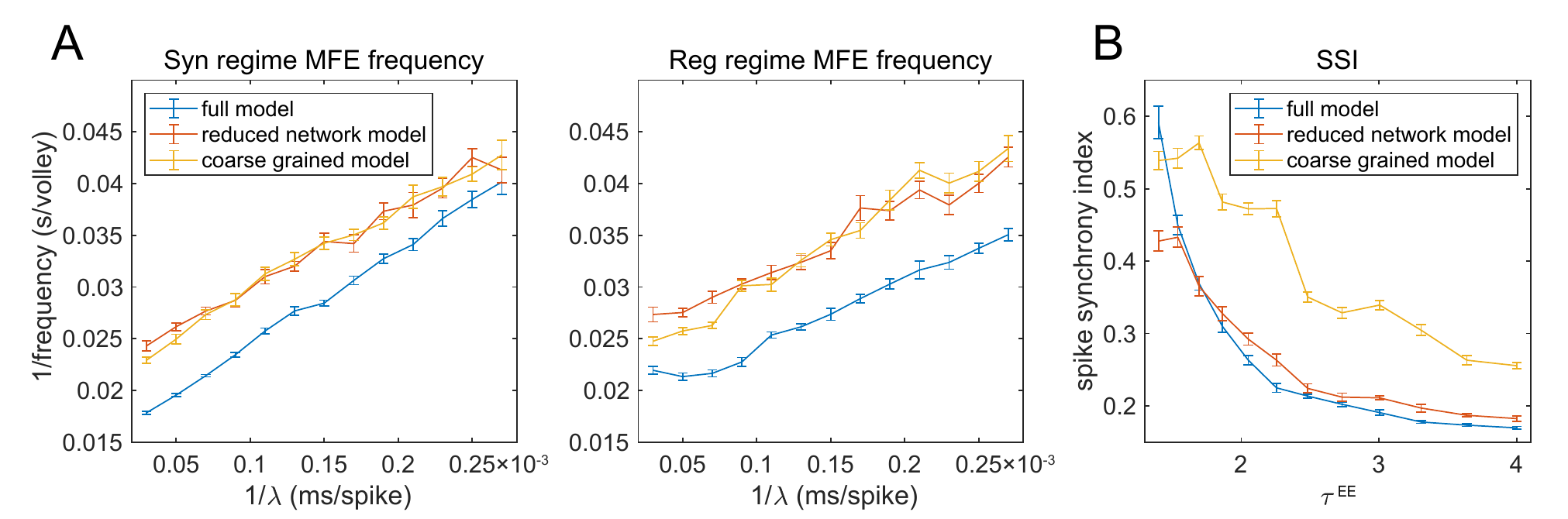}
    \caption{Two gamma features captured by reduced models.  \textbf{A.} MFE waiting time linearly related to external stimulation waiting time. Left: Syn regime; Right: Reg regime. \textbf{B.} Degree of synchrony decreases when $\tau^{EE}$ increases. } 
    \label{Fig5: Gamma Features}
  \end{center}
\end{figure}

The reduced models are not designed to reproduce every detail of the full model. But how well can our model reduction capture the dynamics and key statistical features? In addition to the firing rates, spectral densities, and spike-time correlations for the three selected regimes, we examine the RN and CG models when we change parameters continuously.  In this section, we test two dynamical features observed in the full, MIF model.  For each parameter set involved, we numerically simulate each of the three models (MIF, RN, and CG), for 10 seconds each, and divide the dynamics into 10 batches. We then show the batch means and standard errors of the statistics collected from the batches (Fig.~\ref{Fig5: Gamma Features}).  %\xtext{@Louis, I need some help here for citations:}

First, when the frequency of the external Poisson stimulus ($\lambda$) increases, MFEs appear more frequently in the firing patterns.  According to \cite{chariker2018rhythm}, a new MFE is initiated by excitatory stimulation or by chance when the inhibition from the last MFE fades. Therefore, stronger external stimuli result in faster initiation of a new MFE. From the firing patterns of our three models, we use a spike cluster detection algorithm \cite{chariker2015emergent} to recognize individual MFEs (see Methods) and examine how their emergence is regulated by external stimulus.  We find that the RN and CG models capture the same trend exhibited in the full model (Fig.~\ref{Fig5: Gamma Features}A), namely, the average waiting time of MFE (1/MFE frequency) is linearly related to the external kicks ($\lambda^{-1}$).  However, while the trend is captured semi-quantitatively, the reduced models exhibit lower MFE frequencies (or higher MFE waiting time in Fig.~\ref{Fig5: Gamma Features}A, red and yellow). On the other hand, we find that the MFEs produced by the RN and CG models have longer duration than the MIF model (See Supplementary Materials). These results suggest that, on average, the reduced models have slower probability flows between states. %This is not very surprising: When capturing the dynamics, each state of the reduced models represent multiple states of the MIF model. Therefore, the probability transfer between the states of the reduced models

Second, when the ratio $\nicefrac{\tau^{EE}}{\tau^I}$ becomes smaller, the $E$-kicks take effect on $E$ neurons relatively faster, and thus, the recurrent network excitation recruits other $E$-spikes on a shorter time scale.  Therefore, the whole network exhibits more synchronized firing patterns. This phenomena has been observed in many previous computational models (see, for instance, \cite{KeeleyEtAl2019}), and also verified by the comparison between the three regimes in this paper.  Here we go beyond these three regime and change $\tau^{EE}$ continuously while fixing $\tau^I$ (Fig.~\ref{Fig5: Gamma Features}B).  In the full model, the degree of synchrony (measured by spike synchrony index, see Methods) exhibits a clear decreasing trend when $\tau^{EE}$ goes up, which is also well-echoed by the RN model.  For the CG model, however, although the same trend is captured, the degree of synchrony is generally higher than in the full model.  This bias is introduced when we merge all $E$-kick pools into one and assume $\nicefrac{H^{EE}}{H^{IE}}$ is a constant: The underestimation of $H^{IE}$ delays the spikes of $I$ neurons and the MFEs are artificially prolonged, leading to higher synchrony.  To verify this point, we carry out a CG model reduction with five variables $(N_{GE},N_{GI},H^{EE},H^{IE},H^{I})$ by keeping separately the pending $E$-kick pools of the $E$ and $I$ populations.  In this five-variable CG model, the degree of synchrony is indeed much closer to the full model (see Supplementary Materials). %\tnote{should we put this in the supplementary material?}\xnote{Agreed}

%%%%%%%%%%%%%%%%%%%%%%%%%%%%%%%%%%%%%%%%%%%%%%%%%%%%%%%%%%% Fig5
%%%%%%%%%%%%%%%%%%%%%%%%%%%%%%%%%%%%%%%%%%%%%%%%%%%%%%%%%%% Fig5

\subsection{Gamma Oscillations Remain Near a Low-Dimensional Manifold}
\begin{figure}
  \begin{center}
    %% \captionsetup{type=figure} 
    \includegraphics[width=\textwidth]{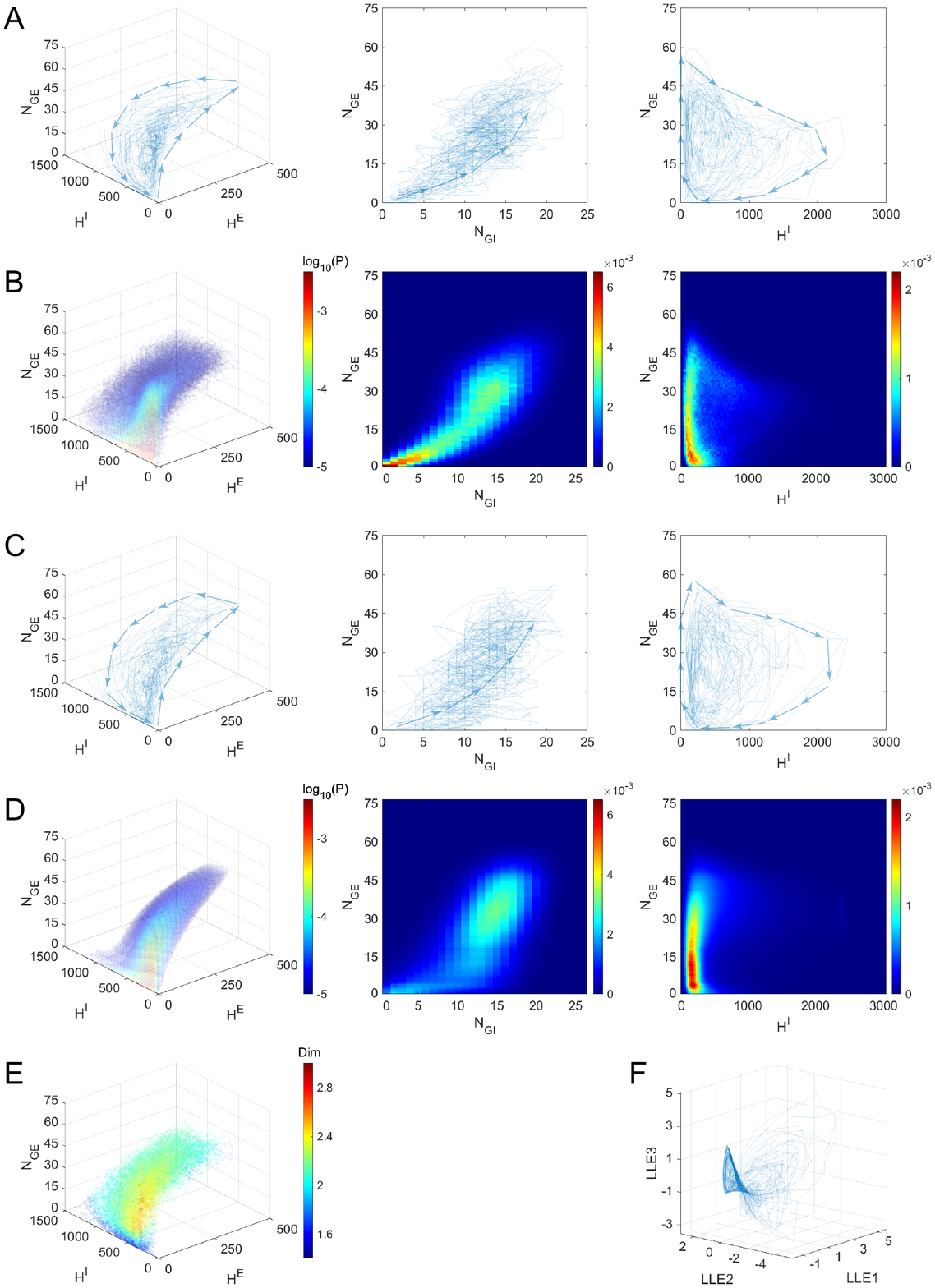}
    \caption{Gamma dynamics restricted on a low-dimensional manifold.  \textbf{A.} The trajectories of the full model (Syn regime) presented the state space $(N_{GE},N_{GI},H^{E},H^{I})$. Left: The projection on subspace $(N_{GE},H^{E},H^{I})$; Middle: The projection on subspace $(N_{GE},N^{GI})$; Right: The projection on subspace $(N_{GE},H^{I})$. \textbf{B-D} depict the same subspaces as \textbf{A}. \textbf{B.} Mass of trajectory density of the full model. \textbf{C.} The trajectories of the CG model (Syn regime). \textbf{D.} The stationary probability distribution of CG model. \textbf{E.} Local dimensionality of data for the full model trajectories, displayed in the same view as \textbf{A}. \textbf{F.} The full model trajectories are local linearly embedded in a three-dimensional space. } 
    \label{Fig6: Manifold}
  \end{center}
\end{figure}

Our most remarkable finding is that the emergent gamma oscillations in the full model stay near a low-dimensional manifold.  Inspired by the fact that the gamma dynamics is successfully captured by the dynamics of only four variables, we simulate the full model in the ``Syn" regime, and then project the trajectories onto the 4-dimensional state space suggested by the success of our CG model, i.e., we collect the statistics $(N_{GE},N_{GI},H^{E},H^{I})$ from the full model dynamics (Fig.~\ref{Fig6: Manifold}A; see also Fig.~\ref{Fig7}A). 

Since $N_{GE}$ and $N_{GI}$ are positively correlated (Fig.~\ref{Fig6: Manifold}A, Middle; See also Fig.~\ref{Fig7}A, bottom panel), we examine the three-dimensional subspace of $(N_{GE},H^{E},H^{I})$. Strikingly, the full model trajectories suggest a low-dimensional dynamical structure of gamma oscillations. This observation is also verified by the mass estimation from the trajectories collected from a long-time (50s) simulation of the full model (Fig.~\ref{Fig6: Manifold}B). On the other hand, we also simulate the CG model for 10 seconds and find similar trajectories in state space (Fig.~\ref{Fig6: Manifold}C), accounting for its successful reproduction of the emergent gamma dynamics. Since the CG model only consists of $O(N^4)$ states, its invariant probability distribution becomes computable from the Markov transition probabilities matrix. Here we present the mass of the distribution after we further ``shrink" the CG model and reduce the number of states to millions (Fig.~\ref{Fig6: Manifold}D. See Methods for the ``shrunk" CG model). The invariant probability distribution not only displays similar mass of density as the full model, but on which the trajectories of the full model remain near (See Supplementary Materials).

The trajectories and the probability distributions of both the full MIF and corresponding CG models reveal a two-dimensional manifold structure with a certain thickness in the orthogonal direction (Fig.~\ref{Fig6: Manifold}A-D).  To verify this point, we carry out a local dimensionality estimation of the full model data (see Methods).  Notably, the manifold exhibits a nearly two-dimensional local structure at most of the places (Fig.~\ref{Fig6: Manifold}E, cyan and green parts), except for a region with low $H^{E},H^{I}$ and low-to-medium $N_{GE}$ (Fig.~\ref{Fig6: Manifold}E, red part. $0<N_{GE}<30$, $H^E<20$, $H^I<200$).  The regions with high dimensionality correspond to the inter-MFE periods (where both number of gate neurons and pending kick pools are small, $0<N_{GE}<15$) and the initiation of MFEs (where $15<N_{GE}<30$).  This is not very surprising: Inter-MFE periods are highly affected by the external stimuli, and the size of each MFE depends on how many $E$-neurons are recurrently recruited by the $E$-spikes at the very beginning.  Both processes are highly stochastic, resulting into higher dimensionalities. We note that, this local two-dimensional structure is also verified by the results of local linear embedding (LLE). When we apply LLE to the 10-second full model trajectories (from four-dimensional to three-dimensional), they display a two-dimensional manifold as well, in the three-dimensional LLE space (Fig.~\ref{Fig6: Manifold}F). 

The trajectories of the MIF model give us a clear view of the temporal organization of the emergent gamma dynamics. First of all, we observed that $(N_{GE},N_{GI})$ are strongly positively correlated (Fig.~\ref{Fig7}A, bottom panel), although $N_{GI}$ rises slightly faster than $N_{GE}$ at the initiation of each MFE. In other words, the probability flow of membrane potential distribution of $E$ and $I$ populations are mostly temporally synchronized, which is also observed in previous computational models \cite{RanganYoung2013a,ZhangRangan2015,chariker2018rhythm}.  In the three-dimensional subspace of $(N_{GE},H^{E},H^{I})$, the low-dimensional manifolds help us interpret the gamma dynamics as follows: 
\begin{enumerate}
    \item At the beginning of each MFE, due to the external stimulus and faded recurrent inhibition, the membrane potentials moves up (represented by increasing $(N_{GE},N_{GI})$. Fig.~\ref{Fig6: Manifold}A-D, Middle). Note that $N_{GI}$ moves up slightly faster due to the lower $I$-to-$I$ connectivity compared to $I$-to-$E$. 
    \item Some $E$ neurons fire, possibly eliciting more spikes from other $E$ neurons, so $H^E$ increases fast during this phase. 
    \item $I$ neurons are excited by $H^E$ at the same time, and $H^I$ increases as well. Note that $H^E$ is always consumed faster due to smaller $\tau^E$. In this phase, $H^E$ attains its peak while $H^I$ still increases rapidly.
    \item $H^E$ is then consumed and the system is now dominated by the inhibition brought by $H^I$ leading to the end of the MFE. High $H^I$ brings down $(N_{GE},N_{GI})$, and terminating the MFE, leading to the inter-MFE-period. The next MFE is unlikely to start until $H^I$ is mostly consumed. 
\end{enumerate}

%%%%%%%%%%%%%%%%%%%%%%%%%%%%%%%%%%%%%%%%%%%%%%%%%%%%%%%%%%% Fig6
%%%%%%%%%%%%%%%%%%%%%%%%%%%%%%%%%%%%%%%%%%%%%%%%%%%%%%%%%%% Fig6

%%%%%%%%%%%%%%%%%%%%%%%%%%%%%%%%%%%%%%%%%%%%%%%%%%%%%%%%%%%%
\section{Discussion\label{sec_discs}}
The vast range of observed neuronal network dynamics presents a tremendous challenge to systems neurophysiologists, data analysts, and computational neuroscientists. Evermore detailed neurophysiological datasets and large-scale neuronal network simulations reveal dynamical interactions on multiple spatial and temporal scales that also participate in essential brain functions. The observed dynamics exhibits rapid, stochastic fluctuations incorporating strong, transient correlations, quite possibly leading to its complexity. Unfortunately, the emergent fluctuating activity cannot effectively be described by standard ensemble averages, as many population methods cannot capture many of the low order statistics associated with the observed dynamical regimes. 

% A challenging issue for theoretical neuroscience is to develop a simple mathematical framework that can effectively reduce and capture a broad range of network dynamics, ranging from homogeneity to synchrony. Here we focus on gamma frequency oscillations because of their role underlying the transient fluctuations within the stationary states of complex neuronal networks (see, for instance, \cite{SiegelDonnerEngel2012,BuzsakiWang2012}), and the belief that they are significant contributors to neural information processing \cite{Fries2009,Wang2010}.

The diversity and hierarchy of experimentally observed dynamics in the brain poses this immediate question: What concise, unified mathematical framework can reproduce the co-existence of transient, heterogeneous dynamical states, emerging from a high-dimensional, strongly recurrent neuronal network. Here we focus on gamma frequency oscillations because of their role underlying the transient fluctuations within the stationary states of complex neuronal networks (see, for instance, \cite{SiegelDonnerEngel2012,BuzsakiWang2012}), and the belief that they are significant contributors to neural information processing \cite{Fries2009,Wang2010}.

Important theoretical progress has been made by coarse graining sparsely coupled networks (see, for instance, \cite{BrunelHakim1999,CaiKinetic2006}, and references therein). These approaches were able to capture spatio-temporal network dynamics dominated by mean-field and uncorrelated synaptic fluctuations. Also influential were the studies focusing on weakly-coupled oscillators. Recent work has made significant theoretical advances by mapping populations of the quadratic IF neurons and the so-called theta neurons to systems of Kuramoto oscillators \cite{MontbrioEtAl2015,Laing2018}, where the tools of phase oscillator theory \cite{ashwin2016mathematical,bick2020understanding} can be used to give insight into network synchronization. We view our approach here as complementary to these studies.

% By using a Markovian IF model first studied by \cite{li2019well}, we show that we can dimensionally reduce and coarse-grain the full model, first, to a two-state reduced network model and then, to a model of transition probabilities between the number of neurons in the two-state, reduced model. The full repertoire of network dynamics between homogeneity and synchrony is faithfully reproduced in our framework.

Here we demonstrate that, by starting with a Markovian IF model, we can bring the tools of classical Markov processes to bear on stochastic gamma oscillations. Using a model first studied by \cite{li2019well}, we show that we can dimensionally reduce and coarse-grain the model network, first, to a two-state reduced network model and then, to a model of transition probabilities between the number of neurons in the two-state, reduced model. The full repertoire of network dynamics between homogeneity and synchrony is faithfully reproduced by our reduced models. Furthermore, by preserving the Markovian dynamics at each step, and combining with data-driven approaches, we were able to reveal locally two-dimesionaly invariant manifolds underlying the type of gamma oscillations observed in large-scale numerical simulations.

A tremendous amount of experimental and theoretical literature implicates oscillatory, coherent neural activity as a crucial element of cognition. Here we provided a simple framework through which collective behavior of populations of neurons can be coarse-grained to counting statistics of neurons in specific neuronal states. This dynamical perspective not only can afford a concise handle for the systems neuroscientists, with experimental access to neuronal circuits and populations, but also to the theoreticians, who may wish to build computational and information processing frameworks on top of these coarse-grained, low-dimensional representations.

While we have detailed our methodology for a system with a small number of neurons and homogeneous connectivities, preliminary studies show that we can scale up to much larger networks (see Supplementary Information). We can extend our framework to networks with slowly varying spatial inhomogeneities (e.g., V1 orientation hypercolumns coupled by long-range excitatory connections) and in capturing interneuronal correlations in predominantly feedforward networks, like synfire chains \cite{DiesmannEtAl1999,WangSornborgerTao,XiaoZhangSornborgerTao}, with each local, nearly homogeneous population described by CG modules (of 2-3 states plus the relevant pending-spike pools). At the same time, we are examining data-driven and machine learning approaches to estimate the various states and transition probabilities directly from numerical simulations, with the hope of applying to neurophysiological data sets. Finally, we are already looking to incorporate higher-order structural motifs in the network connectivity. Higher-order motifs \cite{SongEtAl2005} are likely to substantially influence the types of dynamical correlations within complex networks, and consequently, activate a multitude of spatio-temporal spiking patterns that may have important consequences for information processing and coding in the mammalian brain \cite{zhao2011synchronization,hu2013motif}.

%%%%%%%%%%%%%%%%%%%%%%%%%%%%%%%%%%%%%%%%%%%%%%%%%%%%%%%%%%%%
\section*{Methods\label{sect_mthd}}
\subsection*{Integrate-and-Fire Network}
Consider an $N$-neuron Integrate-and-Fire (IF) neuronal network with $N_E$ excitatory neurons ($E$) and $N_I$ inhibitory neurons ($I$), where the membrane potential ($v_i$) of each neuron is driven by a sum of synaptic currents:
\begin{equation}
\label{sect_mthd:IF}
    \begin{split}
    \ddt{v_i} &= \left(g_i^{ext}+g_i^E\right)\cdot\left(V^E-v_i\right)+g_i^I\left(V^I-v_i\right), \\
    g_i^{ext} &= S_i^{ext}\sum_{\mu_i^{ext}}G^E(t-t_{\mu_i^{ext}}), \\
    g_i^E &= \sum_{\substack{j\in E\\j\neq i}}S_{ij}^{E}\sum_{\mu_j^{E}}G^E(t-t_{\mu_{j}^{E}}), \qquad
    g_i^I = \sum_{\substack{j\in I\\j\neq i}}S_{ij}^{I}\sum_{\mu_j^{I}}G^I(t-t_{\mu_{j}^{I}}),
    \end{split}
\end{equation}
where $g_i^{\{ext,E,I\}}$ are the external, excitatory and inhibitory conductances of neuron $i$.  Each neuron receives excitatory spiking stimulus from an external source ($\mu_i^{ext}$) and other excitatory/inhibitory neurons in the network $\mu_{j}^{\{E,I\}}$, where the strength of synaptic couplings are represented by $S_i^{ext}$ and $S_{ij}^{\{E,I\}}$, respectively.  A spike is released by neuron $i$ when its membrane potential $v_i$ reaches the threshold $V^{th}$. After this, neuron $i$ immediately enters the refractory period, and remains there for a fixed time of $\tau^{\cR}$ before resetting to rest $V^r$.  It is conventional in many previous studies to choose $V^{th} = 1$ and $V^r = 0$ \cite{CaiKinetic2006}.  Accordingly, $V^E = \nicefrac{14}{3}$ and $V^I = -\nicefrac{2}{3}$ are the excitatory and inhibitory reversal potentials.  Each spike changes the postsynaptic conductance with a Green's function,
\begin{equation}
\label{sect_mthd:Spike_Eff}
    \begin{split}
        G^E(t) &= \frac{1}{\tau^E}e^{-\nicefrac{t}{\tau^E}}h(t), \\
        G^I(t) &= \frac{1}{\tau^I}e^{-\nicefrac{t}{\tau^I}}h(t),
    \end{split}
\end{equation}
where $h(t)$ is the Heaviside function.  The time constants, $\tau^{\{E,I\}}$, model the time scale of conductances of the excitatory and inhibitory synapses (such as AMPA and GABA \cite{gerstner2014neuronal}). 

While Eq.~(\ref{sect_mthd:IF}) can model a network with arbitrary connectivity structure, in this paper, we focus on homogeneous networks. That is to say, whether certain spike released by a neuron of type $Q$ is received by another neuron of type $Q'$ is only determined by an independent coin flip with a probability $P_{Q'Q}$, where $Q,Q' \in \{E,I\}$. Furthermore, $S_{ij}^{\{E,I\}}$ are also considered as constants independent of $i,j$. 

Three different levels of models are illustrated below. The \textit{Markovian integrate-and-fire network} approximates Eq.~(\ref{sect_mthd:IF}) with a Markov process, and the following \textit{reduced network} and \textit{coarse-grained model} are reductions of the full Markovian model. 

\subsection*{Full Model: A Markovian Integrate-and-Fire Network}
Following a previous study \cite{li2019well}, we rewrite Eq.~(\ref{sect_mthd:IF}) as a Markov process to facilitate theoretical analysis.  Therefore, we need to minimize the effects of the memory terms and discretize membrane potentials and conductances. Specifically, $v_i$ takes values in 
\begin{equation}
    \label{sect_mthd:MIF}
    \Gamma:=\left\{V^{I},V^{I}+1, \ldots,V^r-1,V^r,V^r+1, \ldots, V^{th}\right\} \cup\{\cR\},
\end{equation}
To be consistent with the IF network~(\ref{sect_mthd:IF}), we choose $V^{I} = -66$, $V^r = 0$, and $V^{th}=100$. As before, $v_i$ enters the refractory state $\cR$ immediately after reaching $V^{th}$.  However, in this Markovian IF (MIF) network, the total time spent in $\cR$ is no longer fixed, but an exponential-distributed random variable $\tau^{\cR}$.% with $\E[\tau^{\cR}] = 2$ms instead. 

Each neuron receives the external input as an independent Poisson process with rate $\lambda_{\{E,I\}}$, and $v_i$ goes up for 1 when an external kick arrives. On the other hand, the synaptic conductances of each neuron $g_i^{\{E,I\}}$ are replaced by ``pending-kick pools," $H_i^{\{E,I\}}$.  Consider excitatory spikes as an example: Instead of updating $g_i^{E}$ with Green's functions when neuron $i$ receives an $E$-spike, we add the new spike to an existing spike pool $H_i^{E}$.  Each spike in the pool will affect $v_i$ independently after an exponentially-distributed waiting time $\tau^{E}$. $H_i^{E}$ is the number of spikes has not taken effect yet.  Therefore, for the a sequence of $E$-spikes received by neuron $i$, it is not hard to see that $\E[H_i^{E}(t)] = \E[g_i^E(t)]$. 

How each spike changes $v_i$ is determined by the type of the spike ($Q'$), the type of neuron $i$ ($Q$), and the state of $v_i$.  When a spike takes effect, the membrane potential stays unchanged if $v_i=\cR$, otherwise $v_i$ may jumps up (for an $E$-spike) or down (for an $I$-spike).  On the other hand, the size of each jump depends on the membrane potential, $v_i$, and the synaptic coupling strengths, $S_{QQ'}$. For an $E$-spike, $v_i$ increases by $S_{QE}$. For an $I$-spike, however, the size of the decrement is $\nicefrac{(v_{i}-V^I)}{(V^{th}-V^I)} \cdot S_{QI}$.  The difference in the effects of $\{E,I\}$-spikes is due to the relative values of the reversal potentials. The currents induced by $g_i^{I}$ is sensitive to $v_i$ while the currents induced by $g_i^{E}$ is much less sensitive, i.e., in Eq.~(\ref{sect_mthd:IF}):
\begin{equation}
    0\leq |v_i-V^I|\leq\frac53, \quad \frac{11}{3}\leq |v_i-V^I|\leq\frac{16}{3}
\end{equation}

In our MIF network, we take most of the system (synaptic and stimulus) parameters directly from \cite{li2019well}, but modified a few to accommodate the smaller network studied here ($N_E = 75$ {\it vs.} $N_E=300\sim1000$ in \cite{li2019well}). The parameters are summarized below:
\begin{itemize}
    \item Frequencies of external input: $\lambda_E = \lambda_I = 7000$ Hz;
    \item Synaptic strength: $S_{EE} = S_{EI} = S_{II} = 20$, and $S_{IE} = 8$;
    \item Probability of spike projections: $P_{EE} = 0.15$, $P_{IE} = P_{EI} = 0.5$, and $P_{II} = 0.4$;
    \item Synaptic time-scales: $\tau^I = 4.5$ ms.  $\tau^E < \tau^I$ to reflect the fact that AMPA is faster than GABA.  We use different choices of $\tau^E$ for regimes implying different dynamics (all indicated in ms): 
    \begin{enumerate}
        \item Homogeneous (``Hom"): $\tau^{EE} = 4$,  $\tau^{IE}= 1.2$,  
        \item Regular (``Reg"): $\tau^{EE} = 1.7$,  $\tau^{IE}= 1.2$, 
        \item Synchronized (``Syn"): $\tau^{EE} = 1.4$,  $\tau^{IE}= 1.2$, 
    \end{enumerate}
\end{itemize}

\subsection*{Reduced Network Consisting of Two-State Neurons}
The \textit{reduced network} (RN) is a direct reduction of the MIF network by reducing the size of the state space for membrane potentials.  In RN, each neuron $i$ is a two-state neuron flipping between ``base" or ``gate" states, i.e., instead of taking values in the state space $\Gamma$, now $v_i\in\Gamma_2 = \{B,G\}$. A neuron in the MIF network is deemed ``base" or ``gate" depending on how likely it is going to fire in the next couple of millisecond: Consider a certain cutoff $V^c\in\Gamma$, neuron $i$ is a gate neuron if $v_i\geq V^c$ since it is closer to the threshold and is only a couple of $E$-kicks away from spiking.  Otherwise, neuron $i$ is a base neuron if $v_i<V^c$ or $v_i = \cR$.  Therefore, a flip from base to gate can take place when an $E$-spike or external stimuli takes effect and $v_i$ crosses $V^c$ from the lower side; on the other hand, a flip from gate to base may be due to $v_i$ crossing $V^c$ from the higher side when 1) an $I$-spike takes effect, or 2) the neuron fires and enters the refractory period.

The network with two-state neurons is reduced from the MIF network by combining states together, but generally we do not expect the full MIF model as a lumpable Markov process \cite{tian2006lumpability}.  Therefore, the appropriate transition probability between the base and gate states should be carefully estimated so that the RN can correctly capture the dynamics of the MIF network.  Since the flip between states can only take place after certain spikes, the possibilities are:
\begin{itemize}
    \item \textbf{Effect of external stimuli}:  When a kick arrives, a base $\{E,I\}$ neuron will become a gate $\{E,I\}$ neuron with probability $\{P_{ex}^{BE},P_{ex}^{BI}\}$, while a gate $\{E,I\}$ neuron will fire and become a base $\{E,I\}$ neuron with probability $\{P_{ex}^{GE},P_{ex}^{GI}\}$.
    \item \textbf{Effects of E-kicks}: Similar types of transitions here but different probabilities due to different sizes of kicks: $\{P_E^{GE},P_E^{GI}, P_E^{BI},P_E^{BE}\}$.
    \item \textbf{Effects of I-kicks}: The I-kicks do not have any effect on a base neuron. But I-kicks will depress a gate neuron to a base neuron with probabilities $\{P_I^{GE}, P_I^{GI}\}$.
\end{itemize}

All transition probabilities listed above are time-dependent and determined by the distribution of membrane potentials of neurons in the network. As a first approximation, we can collect their statistics from a long-time simulation of the MIF network.  Here we illustrate how to compute $P_E^{BE}$ for example, and everything else follows.  Consider the distribution of membrane potentials of $E$-neurons, $p_E(v)$.  Then $P_E^{BE}$ is the conditional probability a base $E$-neuron goes across $V^c$ within one $E$-kick, which is expressed as:
\begin{equation}
    P_E^{BE}\left\{p_E\right\} = \frac{\int_{V^c-S_{EE}}^{V^c}p_E(v) \,\mbox{d}v}{\int_{v<V^c,v = \cR} p_E(v) \,\mbox{d}v}
\end{equation}
However, in RN, we do not see the exact distribution $p_E(v)$, but only the number of base and gate $E$-neurons ($N_{GE},N_{BE}$) instead.  Therefore, to set a closure condition for RN, we consider $P_E^{BE}$ as a function of $N_{BE}$ regardless of the specific distributions, i.e.,
\begin{equation}
    \label{sect_mthd: Two-State-P}
    \overline{P_E^{BE}}(N_{BE}) = \expect*{P_E^{BE}\left\{p_E\right\} |  \int_{v<V^c,v = \cR} p_E(v) \,\mbox{d}v = \frac{N_{BE}}{N_{E}}}
\end{equation}
Finally, to collect $\overline{P_E^{BE}}(N_{BE})$, we run the MIF network simulations for a long time and collect all the events when a $E$-kick takes effect and the membrane potential distribution satisfies the condition listed above.  The estimate of $\overline{P_E^{BE}}(N_{BE})$ is hence the probability of one base $E$-neuron crossing $V^c$ conditioned on these events. 

Readers should note that, the three regimes (Hom, Reg, and Syn) investigated in this paper are only differentiated in the waiting time of kicks, i.e., the transition probabilities induced by single kicks are similar, given the observation that the subthreshold distributions in these regimes are alike. Therefore, to carry out reduction in these regimes, we only need the simulation of one canonical parameter set (say Syn) rather than all of them.   

\subsection*{A Coarse-Grained Approximation}
A \textit{coarse-grained} (CG) approximation is developed to further reduce the number of states of the network.  The MIF network has $O((168\cdot n_H^2)^N)$ states, where $n_H$ is the largest possible number of pending kicks for a neuron.  The number for the RN is lower, as $O((2\cdot n_H^2)^N)$, and yet it still grows exponentially with the size of the network.  This number is already astronomical for the 100-neuron network studied in this paper.

\begin{figure}
  \begin{center}
    %% \captionsetup{type=figure} 
    \includegraphics[width=\textwidth]{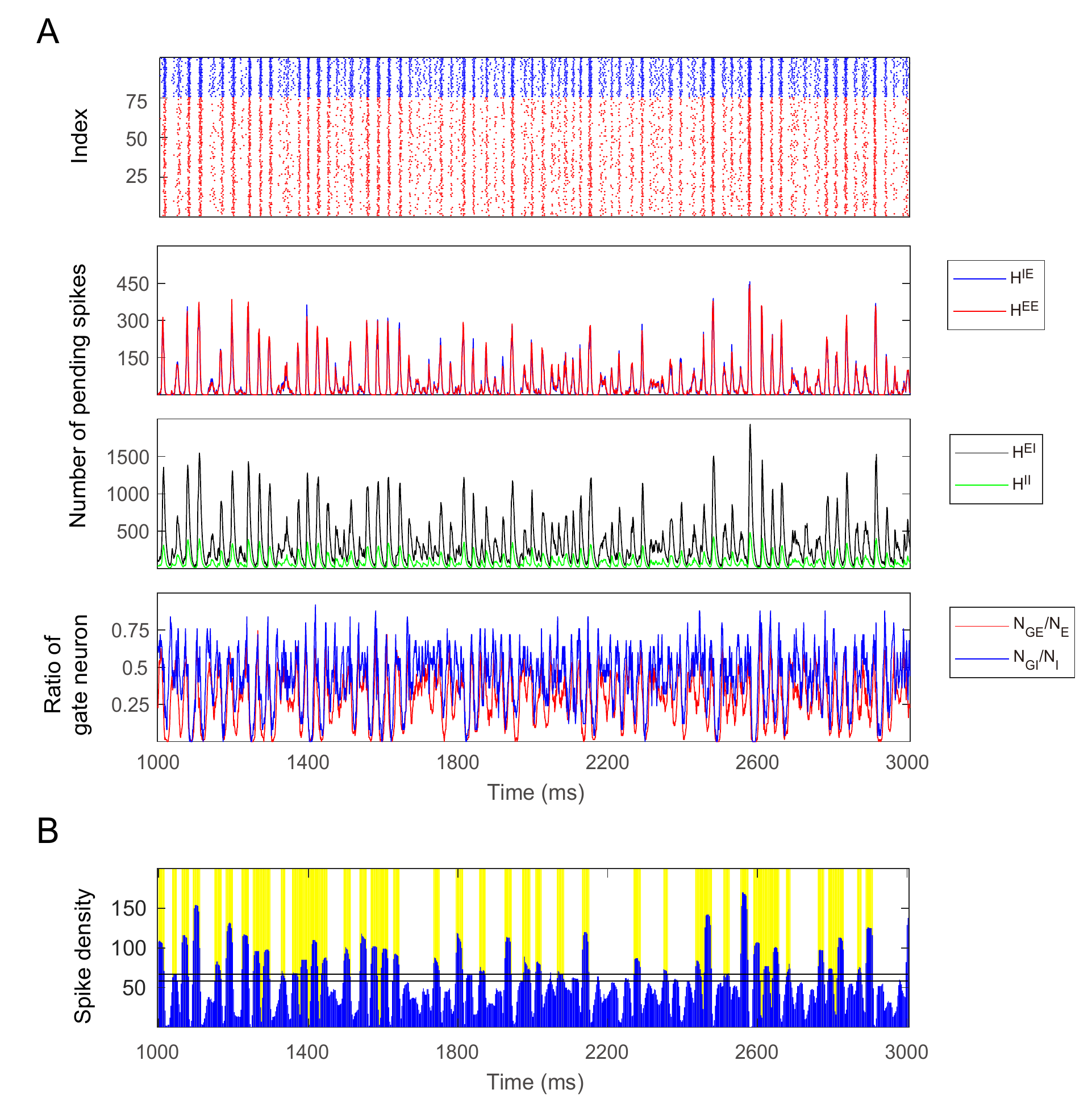}
    \caption{A 2s simulation of the full model in the Syn regime. \textbf{A.} Raster plot (top) and state space statistics (the other three panels).  \textbf{B.} MFEs are recognized by the spiking volley detection algorithm.} 
    \label{Fig7}
  \end{center}
\end{figure}

A CG approximation of RN model is carried out as follows: 
\begin{itemize}
    \item First of all, due to the homogeneous connectivity of the network, all $I$-neurons have the same probability to add an $E$-kick to their pending-kick pools when an $E$-neuron $i$ fires.  Since each kick takes effect independently, this is equivalent to a large pool containing all $E$-kicks of the $I$-neurons and each $E$-kicks are randomly distributed to a specitic $I$-neuron. Therefore, we only need the size of the large pool $H^{IE}$ rather than individual $E$-pending-kick pools for each $I$ neuron.  Likewise, we have pools $H^{EE}$, $H^{EI}$, and $H^{IE}$ for other pending spikes. (Note that this CG simplification does not rule out autapses, i.e., the possibility that a spike takes effect on the neuron releasing it. This may be an issue when the network is very small; however, it does not cause any obvious problems in our model with 100 neurons.)
    
    \item We now try to further combine the pools. Since all $I$-kicks are consumed with the same waiting time $\tau^{I}$, we can combine the two $I$-kick pools together. This does not directly apply to the two $E$-kick pools since $\tau^{EE}\neq\tau^{IE}$, and bias is introduced when we combine $H^{EE}$ and $H^{IE}$, since they are consumed at different rates. We have to assume two constants $a^{EE}+a^{IE} = 1$, where
   \begin{equation}
        H^{EE} = a^{EE}\cdot H^E, \quad H^{IE} = a^{IE}\cdot H^E,
    \end{equation}
    On the other hand, $\tau^{EE}$ and $\tau^{IE}$ are closer in the Syn regime, i.e., $\nicefrac{H^{EE}}{H^{IE}}\approx\mbox{const}$ and the introduced bias is smaller in this regime (Fig.~\ref{Fig7}A). In all, 
    \begin{equation}
        H^E = H^{EE}+H^{IE} = \sum_{i = 1}^N H_i^{E}, \quad H^I = H^{EI}+H^{II} = \sum_{i = 1}^N H_i^{I},
    \end{equation}
    where $H^{\{E,I\}}$ are the sizes of the $\{E,I\}$-kick pools for the whole network.
    
    \item Finally, by definition, the transition probabilities of the two-state neurons are functions of the number of gate neurons  (see Eq.~\ref{sect_mthd: Two-State-P}).  Therefore, instead of the states of each individual neuron, the distributions of membrane potentials can be determined by the numbers of gate neurons $(N_{GE},N_{GI})$.  Once they are computed, the number of base neurons is given by
    \begin{equation}
        N_{GE} + N_{BE} = N_{E},\quad N_{GI}+N_{BI} = N_{I}
    \end{equation}
    \end{itemize}
    
Therefore, the CG approximation above is a Markov process with only four variables, two for the number of gate neurons and two for pending kicks $(N_{GE},N_{GI},H^E,H^I)$, and the number of states is $O(N^4) = N_E\cdot N_I\cdot (n_H)^2$.  This is a tremendous reduction from exponential to polynomial scaling in the size of the network. We here provide a qualitative description of the dynamics of the coarse-grained model: When an $\{E,I\}$-kick takes effect, the number of $H^{\{E,I\}}$ minus one, and certain neuron flips between base/gate states with probability (represented by the change of $N_{GE},N_{GI}$); If an $\{E,I\}$-neuron fires, a spike is released and the pending-kick pool $H^{\{E,I\}}$ expands by 
\begin{equation}
    M_E = P_{EE}N_E+P_{IE}N_I, \quad M_I = P_{EI}N_E+P_{II}N_I,
\end{equation}
where $M_{\{E,I\}}$ is the average number of postsynaptic neuron recipients of an $\{E,I\}$-spike.

We list all possible transitions from state $X = (N_{GE},N_{GI},H^E,H^I)$ (Table 1). In addition, since all state transitions are triggered by certain kicks, which take effect independently with exponential waiting time, it is more important to know the the transition rates based on the transition probabilities. We list these rates in Table 2.
\begin{table}[htbp]
    \centering
    \begin{tabular}{|l|l|l|}
    \hline
    external kick takes effect & one $E$-kick takes effect & one $I$-kick takes effect \\ \hline
    $N_{GE}+1$                 & $N_{GE}+1\quad H^E-1$     &  $N_{GE}-1\quad H^I-1$                       \\ \hline
    $N_{GI}+1$                 & $N_{GI}+1\quad H^E-1$     &  $N_{GI}-1\quad H^I-1$                      \\ \hline
    $N_{GE}-1\quad H^E+M_E$    &$N_{GE}-1 \quad H^E-1+M_E$ & $H^I-1$ \\ \hline
    $N_{GI}-1\quad H^I+S_I$    &$N_{GI}-1 \quad H^E-1\quad H^I+S_I$                        &                      \\ \hline
    Remain                     &$H^E-1$                    &\\ \hline
    \end{tabular}
    \caption{All possible transitions from state $X = (N_{GE},N_{GI},H^E,H^I)$ to another.  There are 13 cases in all.}
    \label{Sect_mthd: table_transitions}
\end{table}
\begin{table}[htbp]
    \centering
    \begin{tabular}{|c|c|c|}
    \hline
external kick takes effect      & one $E$-kick takes effect                                  & one $I$-kick takes effect \\ \hline
$P^{BE}_{ex} N_{BE}\cdot\lambda_E$  & $P^{BE}_{E} a_{EE} \frac{N_{BE}}{N_E} H^E /\tau^{EE}$      & $P^{GE}_{I} a_{EI} \frac{N_{GE}}{N_E} H^I /\tau^{I}$ \\ \hline
$P^{BI}_{ex} N_{BI}\cdot\lambda_I$  & $P^{BI}_{E} a_{IE} \frac{N_{BI}}{N_I} H^E/\tau^{IE}$       & $P^{GI}_{I} a_{II} \frac{N_{GI}}{N_I} H^I /\tau^{I}$ \\ \hline
$P^{GE}_{ex} N_{GE}\cdot\lambda_E$  & $P^{GE}_{E} a_{EE} \frac{N_{GE}}{N_E} H^E/\tau^{EE}$       & $(1- P^{GE}_{I} a_{EI} \frac{N_{GE}}{N_E} -P^{GI}_{I} a_{II} \frac{N_{GI}}{N_I})\cdot  H^I/\tau^I$ \\ \hline
$P^{GI}_{ex} N_{GI}\cdot\lambda_I$  & $P^{GI}_{E} a_{IE} \frac{N_{GI}}{N_I} H^E/\tau^{IE}$       & \\ \hline
$\begin{aligned} &(1-P^{BE}_{ex}) N_{BE} \cdot\lambda_{E}\\+&(1-P^{BI}_{ex}) N_{BI} \cdot\lambda_{I} \\ +&(1-P^{GE}_{ex}) N_{GE} \cdot\lambda_{E} \\ +& (1-P^{GI}_{ex}) N_{GI} \cdot\lambda_{I} \end{aligned}$ & 
%\tabincell{c}{$(1-P^{BE}_{ex}) N_{BE} \cdot\lambda_{E}$\\$+(1-P^{BI}_{ex}) N_{BI} \cdot\lambda_{I}$ \\ $+(1-P^{GE}_{ex}) N_{GE} \cdot\lambda_{E}$\\$+(1-P^{GI}_{ex}) N_{GI} \cdot\lambda_{I} $  }
$\begin{aligned} &(1-P^{BE}_{E}) a_{EE} \frac{N_{BE}}{N_E} H^E /\tau^{EE} \\ +&(1-P^{BI}_{E}) a_{IE} \frac{N_{BI}}{N_I} H^E/\tau^{IE} \\ +&(1-P^{GE}_{E}) a_{EE} \frac{N_{GE}}{N_E} H^E/\tau^{EE} \\ +&(1-P^{GI}_{E}) a_{IE} \frac{N_{GI}}{N_I} H^E/\tau^{IE} \end{aligned}$
%& \tabincell{c}{$(1-P^{BE}_{E}) a_{EE} \frac{N_{BE}}{N_E} H^E /\tau^{EE}$\\$+(1-P^{BI}_{E}) a_{IE} \frac{N_{BI}}{N_I} H^E/\tau^{IE}$ \\ $+(1-P^{GE}_{E}) a_{EE} \frac{N_{GE}}{N_E} H^E/\tau^{EE}$\\$ + (1-P^{GI}_{E}) a_{IE} \frac{N_{GI}}{N_I} H^E/\tau^{IE} $  }
&\\ \hline
\end{tabular}
    \caption{The transition rates of all transitions in Table \ref{Sect_mthd: table_transitions}.}
    \label{Sect_mthd: table_transitions_rates}
\end{table}

%%%%%%%%%%%%%%%%%%%%%%%%%%%%%%%%%%%%%%%%%%%%%%%%%%%%%%%%%%% Fig7
%%%%%%%%%%%%%%%%%%%%%%%%%%%%%%%%%%%%%%%%%%%%%%%%%%%%%%%%%%% Fig7

\subsection*{Statistics}
To quantify how well the reduced models (RN and CG) capture the dynamical features of the full model (MIF), we compare several statistics of the network dynamics (or more precisely, the spiking pattern produced by the network) collected from the simulations of the MIF, RN, and CG models.  The reader should note that we can not tell the specific neuron indices of firing events in the CG model; yet it does not affect the computation of the statistics below. The raster plots produced for the CG model, however, are indeed mock-up raster plots, drawn by assigning spikes to neurons randomly among the appropriate $\{E,I\}$ population. 

\heading{Firing rates.} Spikes from $\{E,I\}$-cells are collected separately, and firing rates $(fr_E,fr_I)$ are computed as the average numbers of spikes per neuron per second.  All three models are simulated simulated for over 3 seconds and spikes are collected from the 2nd second to rule out possible influences by the choice of initial conditions.

\heading{Spike synchrony index.} We borrow the definition of spike synchrony index (SSI) from \cite{chariker2018rhythm}. SSI describes the degree of synchrony of the firing events as the following. For each spike occurred at $t$, consider a $w$-ms time window centered by the spike $(t-\nicefrac{w}{2},t+\nicefrac{w}{2})$ and count the fraction of neurons in the whole network firing in such window.  Finally, the SSI is the fraction averaged over all spikes.

It is not hard to see that SSI is larger for more synchronous spiking patterns. For the completely synchronized dynamics, every other neuron fires within the time window of each spike hence SSI=1. For completely uncorrelated firing patterns such as Poisson, SSI is a small number close to 0. One should note that the absolute value of SSI dependes on the choice of the window size, and we choose $w=5$ms (the same as \cite{chariker2018rhythm}).

\heading{Spectrogram.} The power spectrum density (PSD) measures the variance in a signal as a function of frequency.  In this study, the PSD is computed as follows:

A time interval $(0,T)$ is divide into time bins $B_n=[(n-1)\Delta t, n \Delta t], n=1,2,...$, the spike density $\mu_n$ per neuron in $B_n$ is given by $\mu_n = \nicefrac{m_n}{N\Delta t}$ where $m_n$ is the total number of spikes fired in bin $B_n$. Hence, the discrete Fourier transform of $\{\mu_n\}$ on $(0,T)$ is given as:  
\begin{equation}
    \hat{\mu}(k) = \frac{1}{\sqrt{T}}\sum_{n=1}^{\nicefrac{T}{\Delta t}}\mu_n\Delta t e^{-k\cdot(2\pi i)\cdot(n\Delta t)}.
\end{equation}
Finally, as a function of $k$, PSD is the ``power" concentrated at frequency $k$, i.e., $|\hat{\mu}(k)|^2$.

\heading{Spike timing correlations.} The correlation diagrams describe the averaged correlation between each spike and others. Consider the correlation with $I$-spikes conditioned on E at t = 0:

For each $E$-spike at time $t$, we take $I$-spikes within the time window $[t-15\mbox{ms},t+15\mbox{ms}]$, and compute the fraction of $I$-spikes in each 1-ms time bin. The correlation diagrams is then averaged over all $E$-spikes in this simulation.

\heading{Spiking volley detection.} The method defining spiking volleys (or MFEs) is borrowed from \cite{chariker2015emergent}. The core idea of this method is to find time intervals with some length constraints that the firing rate of each time bin in this interval is a certain amount higher than the average firing rate. This is then defined as a spiking volley. We choose 1 ms time bin and $\delta=0.33,\ \epsilon=8$ as parameters (Fig.~\ref{Fig7}B). 

\subsection*{Computational Methods}
\heading{Exact timing of events.} During our simulation, we can compute the exact timing of all events including firing, kicks taking effect, etc. We note that MIF, RN, and CG models are all Markov processes whose randomness is mainly due to the exponential distributions of various waiting times. Consider two independent events $A$ and $B$ with waiting time $X_A\sim \mbox{exp}(\lambda_A)$, $X_B\sim \mbox{exp}(\lambda_B)$, we have $\min\{X_A, X_B\} \sim \mbox{exp}(\lambda_A+ \lambda_B)$, i.e., the waiting time for either the first event is also an exponential distribution. Furthermore, the probability for the first occurring events to be $A$ is $\nicefrac{\lambda_A}{(\lambda_A+ \lambda_B)}$. 

Similar arguments extend to $m$ events. By noticing that exponential distributed waiting times are temporally memoryless, we can simulated all three models by repeatedly selecting the first occurring event and generate the actual waiting time by sampling from certain exponential distributions. 

\heading{Invariant probability distributions.} After determining the total number of states and the transition probabilities between them, we can calculate the invariant probability distribution from the CG model by computing the eigenvectors and eigenvalues of the transition probability matrix. Such methods can be found in standard linear algebra textbooks such as \cite{roman2005advanced}.

Readers should note that theoretically, there is no upper bound for $\{H^E,H^I\}$. In order to close the computation of invariant probability distributions, we set the $n_{\{H^E,H^I\}}$, the largest numbers of pending spikes shown up in the simulations as the ``boundaries" of the state space. Specifically, the transition probability from state $X$ to $Y$ is 0 if 
\begin{enumerate}
    \item $X = (N_{GE},N_{GI},n_{H^E},H^{I})$ and $Y = (N_{GE},N_{GI},n_{H^E}+a,H^{I})$, or
    \item $X = (N_{GE},N_{GI},H^E,n_{H^{I}})$ and $Y = (N_{GE},N_{GI},H^E,n_{H^{I}}+b)$,
\end{enumerate}
where $a,b>0$.

\heading{A ``Shrunk" Coarse Grained Model.} After the reductions, the CG model studied in this paper still has $M = 5.6\times10^9$ states.  Though the first left eigenvector (i.e., the stationary probability distribution, corresponding to eigenvalue 1) of a sparse, $M$-by-$M$ probability transition matrix is computable, the cost could be high for a desktop. Therefore, we aim at an even coarser version of the CG model: a ``shrunk" coarse grained (SCG) model. Since $H^{\{E,I\}}$ are much higher than $N_{\{GE,GI\}}$, we shrink the CG model by combining every $K$ states of pending kicks into one, i.e., all $1\leq H^{E}_{\mbox{cg}}\leq K$ states in CG model is considered as $H^{E}_{\mbox{scg}}=1$ in SCG model. The intuition is the following: no need to characterize the states of pending spikes very precisely especially when the number is very large (i.e., the difference between 3000 and 3001 is very small). We choose $K=24$ to keep it the same as $S_{EE}$.

Every state of SCG model can also be represented as a quadruplet $\mathbb{Q}_{\mbox{scg}}=(N_{GE},N_{GI},H^{E}_{\mbox{scg}},H^{I}_{\mbox{scg}})$. The SCG model works as follow. Firstly the quadruple $\mathbb{Q}_{\mbox{scg}}$ is lifted to another quadruplet $\mathbb{Q}_{\mbox{cg}}=(N_{GE},N_{GI},H^{E}_{\mbox{cg}},H^{I}_{\mbox{cg}})$, where $H^{E,I}_{\mbox{cg}}=(H^{E,I}_{\mbox{scg}}-0.5)\cdot K$. Then quadruplet $\mathbb{Q}_{\mbox{cg}}$ acts following the same rule as the CG model. Lastly the change of $\mathbb{Q}_{\mbox{cg}}$ is projected back into the change of $\mathbb{Q}_{\mbox{scg}}$:
\begin{enumerate}
    \item The change of $N_{GE}$ and $N_{GE}$ in $\mathbb{Q}_{\mbox{cg}}$ is kept the same on corresponding elements in $\mathbb{Q}_{\mbox{scg}}$;
    \item The change $x$ of $H^{E,I}_{\mbox{cg}}$ is replaced by change $y=[x/K]+b$ on $H^{E,I}_{\mbox{scg}}$, where $[\cdot]$ denotes the least integer function and $b$ is a Bernoulli variable with probability $p=(x/K)-y$;
\end{enumerate}

Through this model, we can further reduce the $M$ states of CG model to $M/K^2$ states of SCG model.

\heading{Locally-linear embedding.} LLE is a nonlinear dimension reduction method which discovers the low-dimensional structure of high-dimensional data \cite{roweis2000nonlinear}. More precisely, it maps the high-dimensional input data into a low-dimensional space. The core idea of LLE is to maintain the local linear structure through the mapping and this is achieved by a two-step optimization. There are $N$ input data and we denote the high-dimensional input as $\Vec{X}_i$ and low-dimensional output as $\Vec{Y}_i$. The algorithm is described below and more details see \cite{roweis2000nonlinear,saul2000introduction}.
\begin{enumerate}
    \item Find the nearest $k$ neighbors $S_i^k$ of each data point $\Vec{X}_i$;
    \item $W = \mathop{\arg\min}_{W^\prime}\sum_i\|\Vec{X}_i-\sum_j W_{ij}^\prime \Vec{X}_j\|^2$,$\ W_{ij}^\prime=0$ if $\Vec{X}_j\notin S_i^k$;
    \item $Y = \mathop{\arg\min}_{Y^\prime}\sum_i\|\Vec{Y}^\prime_i-\sum_j W_{ij}\Vec{Y}^\prime_j\|^2$, subject to $\sum_i\Vec{Y}^\prime_i =0$  and  $\frac{1}{N}\sum_i\Vec{Y}^\prime_i\Vec{Y}_i^{\prime T}=I $;
\end{enumerate}

\heading{Dimensionality of data.} We use local principal component analysis (local PCA) method to compute the local dimensionality of the data at each data point $x$, i.e., how its neighbors cluster around $x$. The data points processed here is selected from original data points with probability proportional to the square of distance from the place with highest density mass, in order to make the distribution of data points more uniform which can result in more precise local dimensionality chracterization. For $x$, consider the correlation matrix $C_x$ of $x$ and its $K$ nearest neighbors ($K$ selected as 100). Then the dimensionality at $x$ is computed as %  \xnote{@Tianyi, is this the reason why the bottom part has low dimension? I suggest we use the distance from the place with highest density mass, instead of the origin (0,0,0,0)}
\begin{equation}
    \label{Methods: Dim}
    \mbox{Dim}(x)=\frac{\mbox{Tr}(C_x)^2}{\mbox{Tr}(C_x^2)}=\frac{(\sum_i\lambda_i)^2}{\sum_i\lambda_i^2}
\end{equation}
where $\lambda_i$ is the $i$-th eigenvalue of correlation matrix $C_x$. Eq.~(\ref{Methods: Dim}) is widely used as the dimensionality definition in theoretical and experimental neuroscience studies \cite{gao2017theory,recanatesi2019dimensionality,mazzucato2016stimuli,litwin2017optimal}.

%%%%%%%%%%%%%%%%%%%%%%%%%%%%%%%%%%%%%%%%%%%%%%%%%%%%%%%%%%%%
\section*{Acknowledgments} This work was partially supported by the Natural Science Foundation of China through grants 31771147 (T.W., L.T.) and 91232715 (L.T.), by the Open Research Fund of the State Key Laboratory of Cognitive Neuroscience and Learning grant CNLZD1404 (T.W., L.T.). Z.X. is supported by the Swartz Foundation through the Swartz Postdoctoral Fellowship.

We thank Professors Yao Li (University of Massachusetts, Amherst), Kevin Lin (University Arizona), and Lai-Sang Young (New York University) for their useful comments.

%% BibTeX users please use one of
%%\bibliographystyle{spbasic}      % basic style, author-year citations
%%\bibliographystyle{spphys}       % APS-like style for physics
%\bibliography{a}   % name your BibTeX data base
%
%
%%%%%%%%%%%%%%%%%%%%%%%%%%%%%%%%%%%%%%%%%%%%%%%%%%%%%%%%%%%%%%%%%%%%%%%%%
%\bibliographystyle{siamplain}

\printbibliography

@Article{WangSornborgerTao,
  title={Graded, dynamically routable information processing with synfire-gated synfire chains},
  author={Wang, Zhuo and Sornborger, Andrew T and Tao, Louis},
  journal={PLoS Computational Biology},
  volume={12},
  number={6},
  pages={e1004979},
  year={2016},
  publisher={Public Library of Science San Francisco, CA USA}
 }

@article{XiaoZhangSornborgerTao,
  title={Cusps enable line attractors for neural computation},
  author={Xiao, Zhuocheng and Zhang, Jiwei and Sornborger, Andrew T and Tao, Louis},
  journal={Physical Review E},
  volume={96},
  number={5},
  pages={052308},
  year={2017},
  publisher={APS}
 }

@article{bauer2007gamma,
  title={Gamma oscillations coordinate amygdalo-rhinal interactions during learning},
  author={Bauer, Elizabeth P and Paz, Rony and Par{\'e}, Denis},
  journal={Journal of Neuroscience},
  volume={27},
  number={35},
  pages={9369--9379},
  year={2007},
  publisher={Soc Neuroscience}
}

@article{beggs2003neuronal,
  title={Neuronal avalanches in neocortical circuits},
  author={Beggs, John M and Plenz, Dietmar},
  journal={Journal of Neuroscience},
  volume={23},
  number={35},
  pages={11167--11177},
  year={2003},
  publisher={Soc Neuroscience}
}

@article{bressler2003context,
  title={Context rules},
  author={Bressler, Steven L},
  journal={Behavioral and Brain Sciences},
  volume={26},
  number={1},
  pages={85--85},
  year={2003},
  publisher={Cambridge University Press}
}

@article{BrunelHakim1999,
  title={Fast global oscillations in networks of integrate-and-fire neurons with low firing rates},
  author={Brunel, Nicolas and Hakim, Vincent},
  journal={Neural Computation},
  volume={11},
  number={7},
  pages={1621--1671},
  year={1999},
  publisher={MIT Press}
}

@article{buice2013beyond,
  title={Beyond mean field theory: statistical field theory for neural networks},
  author={Buice, Michael A and Chow, Carson C},
  journal={Journal of Statistical Mechanics: Theory and Experiment},
  volume={2013},
  number={03},
  pages={P03003},
  year={2013},
  publisher={IOP Publishing}
}

@article{CaiKinetic2006,
  title={Kinetic theory for neuronal network dynamics},
  author={Cai, David and Tao, Louis and Rangan, Aaditya V and McLaughlin, David W and others},
  journal={Communications in Mathematical Sciences},
  volume={4},
  number={1},
  pages={97--127},
  year={2006},
  publisher={International Press of Boston}
}

@article{chariker2015emergent,
  title={Emergent spike patterns in neuronal populations},
  author={Chariker, Logan and Young, Lai-Sang},
  journal={Journal of Computational Neuroscience},
  volume={38},
  number={1},
  pages={203--220},
  year={2015},
  publisher={Springer}
}

@article{chariker2018rhythm,
  title={Rhythm and synchrony in a cortical network model},
  author={Chariker, Logan and Shapley, Robert and Young, Lai-Sang},
  journal={Journal of Neuroscience},
  volume={38},
  number={40},
  pages={8621--8634},
  year={2018},
  publisher={Soc Neuroscience}
}

@article{churchland2010stimulus,
  title={Stimulus onset quenches neural variability: a widespread cortical phenomenon},
  author={Churchland, Mark M and Byron, M Yu and Cunningham, John P and Sugrue, Leo P and Cohen, Marlene R and Corrado, Greg S and Newsome, William T and Clark, Andrew M and Hosseini, Paymon and Scott, Benjamin B and others},
  journal={Nature Neuroscience},
  volume={13},
  number={3},
  pages={369--378},
  year={2010},
  publisher={Nature Publishing Group}
}

@article{colgin2016rhythms,
  title={Rhythms of the hippocampal network},
  author={Colgin, Laura Lee},
  journal={Nature Reviews Neuroscience},
  volume={17},
  number={4},
  pages={239--249},
  year={2016},
  publisher={Nature Publishing Group}
}

@inproceedings{cowan1991stochastic,
  title={Stochastic neurodynamics},
  author={Cowan, Jack D},
  booktitle={Advances in Neural Information Processing Systems},
  pages={62--69},
  year={1991}
}

@article{gao2017theory,
  title={A theory of multineuronal dimensionality, dynamics and measurement},
  author={Gao, Peiran and Trautmann, Eric and Yu, Byron and Santhanam, Gopal and Ryu, Stephen and Shenoy, Krishna and Ganguli, Surya},
  journal={BioRxiv},
  pages={214262},
  year={2017},
  publisher={Cold Spring Harbor Laboratory}
}

@book{gerstner2014neuronal,
  title={Neuronal Dynamics: From Single Neurons to Networks and Models of Cognition},
  author={Gerstner, Wulfram and Kistler, Werner M and Naud, Richard and Paninski, Liam},
  year={2014},
  publisher={Cambridge University Press}
}

@article{goddard2012gamma,
  title={Gamma oscillations are generated locally in an attention-related midbrain network},
  author={Goddard, C Alex and Sridharan, Devarajan and Huguenard, John R and Knudsen, Eric I},
  journal={Neuron},
  volume={73},
  number={3},
  pages={567--580},
  year={2012},
  publisher={Elsevier}
}

@article{hodgkin1952quantitative,
  title={A quantitative description of membrane current and its application to conduction and excitation in nerve},
  author={Hodgkin, Alan L and Huxley, Andrew F},
  journal={The Journal of Physiology},
  volume={117},
  number={4},
  pages={500},
  year={1952},
  publisher={Wiley-Blackwell}
}

@article{hu2013motif,
  title={Motif statistics and spike correlations in neuronal networks},
  author={Hu, Yu and Trousdale, James and Josi{\'c}, Kre{\v{s}}imir and Shea-Brown, Eric},
  journal={Journal of Statistical Mechanics: Theory and Experiment},
  volume={2013},
  number={03},
  pages={P03012},
  year={2013},
  publisher={IOP Publishing}
}

@article{krystal2017impaired,
  title={Impaired tuning of neural ensembles and the pathophysiology of schizophrenia: a translational and computational neuroscience perspective},
  author={Krystal, John H and Anticevic, Alan and Yang, Genevieve J and Dragoi, George and Driesen, Naomi R and Wang, Xiao-Jing and Murray, John D},
  journal={Biological Psychiatry},
  volume={81},
  number={10},
  pages={874--885},
  year={2017},
  publisher={Elsevier}
}

@article{lee2003synchronous,
  title={Synchronous gamma activity: a review and contribution to an integrative neuroscience model of schizophrenia},
  author={Lee, Kwang-Hyuk and Williams, Leanne M and Breakspear, Michael and Gordon, Evian},
  journal={Brain Research Reviews},
  volume={41},
  number={1},
  pages={57--78},
  year={2003},
  publisher={Elsevier}
}

@article{li2019well,
  title={How well do reduced models capture the dynamics in models of interacting neurons?},
  author={Li, Yao and Chariker, Logan and Young, Lai-Sang},
  journal={Journal of Mathematical Biology},
  volume={78},
  number={1-2},
  pages={83--115},
  year={2019},
  publisher={Springer}
}

@article{li2019stochastic,
  title={Stochastic neural field model: multiple firing events and correlations},
  author={Li, Yao and Xu, Hui},
  journal={Journal of Mathematical Biology},
  volume={79},
  number={4},
  pages={1169--1204},
  year={2019},
  publisher={Springer}
}

@article{litwin2017optimal,
  title={Optimal degrees of synaptic connectivity},
  author={Litwin-Kumar, Ashok and Harris, Kameron Decker and Axel, Richard and Sompolinsky, Haim and Abbott, LF},
  journal={Neuron},
  volume={93},
  number={5},
  pages={1153--1164},
  year={2017},
  publisher={Elsevier}
}

@article{mably2018gamma,
  title={Gamma oscillations in cognitive disorders},
  author={Mably, Alexandra J and Colgin, Laura Lee},
  journal={Current Opinion in Neurobiology},
  volume={52},
  pages={182--187},
  year={2018},
  publisher={Elsevier}
}

@article{mazzucato2016stimuli,
  title={Stimuli reduce the dimensionality of cortical activity},
  author={Mazzucato, Luca and Fontanini, Alfredo and La Camera, Giancarlo},
  journal={Frontiers in Systems Neuroscience},
  volume={10},
  pages={11},
  year={2016},
  publisher={Frontiers}
}

@article{menon1996spatio,
  title={Spatio-temporal correlations in human gamma band electrocorticograms},
  author={Menon, V and Freeman, WJ and Cutillo, BA and Desmond, JE and Ward, MF and Bressler, SL and Laxer, KD and Barbaro, N and Gevins, AS},
  journal={Electroencephalography and Clinical Neurophysiology},
  volume={98},
  number={2},
  pages={89--102},
  year={1996},
  publisher={Elsevier}
}

@article{mcnally2016gamma,
  title={Gamma band oscillations: a key to understanding schizophrenia symptoms and neural circuit abnormalities},
  author={McNally, James M and McCarley, Robert W},
  journal={Current Opinion in Psychiatry},
  volume={29},
  number={3},
  pages={202},
  year={2016},
  publisher={NIH Public Access}
}

@article{MontbrioEtAl2015,
  title={Macroscopic description for networks of spiking neurons},
  author={Montbri{\'o}, Ernest and Paz{\'o}, Diego and Roxin, Alex},
  journal={Physical Review X},
  volume={5},
  number={2},
  pages={021028},
  year={2015},
  publisher={APS}
}

@article{plenz2011multi,
  title={Multi-electrode array recordings of neuronal avalanches in organotypic cultures},
  author={Plenz, Dietmar and Stewart, Craig V and Shew, Woodrow and Yang, Hongdian and Klaus, Andreas and Bellay, Tim},
  journal={JoVE (Journal of Visualized Experiments)},
  number={54},
  pages={e2949},
  year={2011}
}

@article{recanatesi2019dimensionality,
  title={Dimensionality in recurrent spiking networks: global trends in activity and local origins in connectivity},
  author={Recanatesi, Stefano and Ocker, Gabriel Koch and Buice, Michael A and Shea-Brown, Eric},
  journal={PLoS Computational Biology},
  volume={15},
  number={7},
  pages={e1006446},
  year={2019},
  publisher={Public Library of Science}
}

@book{roman2005advanced,
  title={Advanced Linear Algebra},
  author={Roman, Steven and Axler, S and Gehring, FW},
  volume={3},
  year={2005},
  publisher={Springer}
}

@article{roweis2000nonlinear,
  title={Nonlinear dimensionality reduction by locally linear embedding},
  author={Roweis, Sam T and Saul, Lawrence K},
  journal={Science},
  volume={290},
  number={5500},
  pages={2323--2326},
  year={2000},
  publisher={American Association for the Advancement of Science}
}

@article{saul2000introduction,
  title={An introduction to locally linear embedding},
  author={Saul, Lawrence K and Roweis, Sam T},
  journal={unpublished. Available at: http://www. cs. toronto. edu/\~{} roweis/lle/publications. html},
  year={2000}
}

@article{shew2011information,
title={Information capacity and transmission are maximized in balanced cortical networks with neuronal avalanches},
author={Shew, Woodrow L and Yang, Hongdian and Yu, Shan and Roy, Rajarshi and Plenz, Dietmar},
journal={Journal of Neuroscience},
volume={31},
number={1},
pages={55--63},
year={2011},
publisher={Soc Neuroscience}
}

@article{tian2006lumpability,
  title={Lumpability and commutativity of Markov processes},
  author={Tian, Jianjun Paul and Kannan, D},
  journal={Stochastic Analysis and Applications},
  volume={24},
  number={3},
  pages={685--702},
  year={2006},
  publisher={Taylor \& Francis}
}

@article{van2010learning,
  title={Learning-associated gamma-band phase-locking of action--outcome selective neurons in orbitofrontal cortex},
  author={van Wingerden, Marijn and Vinck, Martin and Lankelma, Jan V and Pennartz, Cyriel MA},
  journal={Journal of Neuroscience},
  volume={30},
  number={30},
  pages={10025--10038},
  year={2010},
  publisher={Soc Neuroscience}
}

@article{wang1996gamma,
  title={Gamma oscillation by synaptic inhibition in a hippocampal interneuronal network model},
  author={Wang, Xiao-Jing and Buzs{\'a}ki, Gy{\"o}rgy},
  journal={Journal of Neuroscience},
  volume={16},
  number={20},
  pages={6402--6413},
  year={1996},
  publisher={Soc Neuroscience}
}

@article{whittington2000inhibition,
  title={Inhibition-based rhythms: experimental and mathematical observations on network dynamics},
  author={Whittington, Miles A and Traub, RD and Kopell, N and Ermentrout, B and Buhl, EH},
  journal={International Journal of Psychophysiology},
  volume={38},
  number={3},
  pages={315--336},
  year={2000},
  publisher={Elsevier}
}

@article{wilson1972excitatory,
  title={Excitatory and inhibitory interactions in localized populations of model neurons},
  author={Wilson, Hugh R and Cowan, Jack D},
  journal={Biophysical Journal},
  volume={12},
  number={1},
  pages={1--24},
  year={1972},
  publisher={Elsevier}
}

@article{wilson1973mathematical,
  title={A mathematical theory of the functional dynamics of cortical and thalamic nervous tissue},
  author={Wilson, Hugh R and Cowan, Jack D},
  journal={Kybernetik},
  volume={13},
  number={2},
  pages={55--80},
  year={1973},
  publisher={Springer}
}

@article{xing2012stochastic,
  title={Stochastic generation of gamma-band activity in primary visual cortex of awake and anesthetized monkeys},
  author={Xing, Dajun and Shen, Yutai and Burns, Samuel and Yeh, Chun-I and Shapley, Robert and Li, Wu},
  journal={Journal of Neuroscience},
  volume={32},
  number={40},
  pages={13873--13880a},
  year={2012},
  publisher={Soc Neuroscience}
}

@article{yu2010membrane,
  title={Membrane potential synchrony in primary visual cortex during sensory stimulation},
  author={Yu, Jianing and Ferster, David},
  journal={Neuron},
  volume={68},
  number={6},
  pages={1187--1201},
  year={2010},
  publisher={Elsevier}
}

@article{yu2011higher,
  title={Higher-order interactions characterized in cortical activity},
  author={Yu, Shan and Yang, Hongdian and Nakahara, Hiroyuki and Santos, Gustavo S and Nikoli{\'c}, Danko and Plenz, Dietmar},
  journal={Journal of Neuroscience},
  volume={31},
  number={48},
  pages={17514--17526},
  year={2011},
  publisher={Soc Neuroscience}
}

@article{zhao2011synchronization,
  title={Synchronization from second order network connectivity statistics},
  author={Zhao, Liqiong and Beverlin, Bryce II and Netoff, Tay and Nykamp, Duane Quinn},
  journal={Frontiers in Computational Neuroscience},
  volume={5},
  pages={28},
  year={2011},
  publisher={Frontiers}
}

@article{GrayEtAl1989,
   author    = {Gray, C.M. and K\"onig, P. and Engel, A.K. and Singer, W.},
   title         = {Oscillatory responses in cat visual cortex exhibit inter-columnar synchronization which reflects global stimulus properties}, 
   year       = {1989},
   journal   = {Nature},
   volume  = {338},
   pages    = {334-337}
}

@article{bragin1995gamma,
  title={Gamma (40-100 Hz) oscillation in the hippocampus of the behaving rat},
  author={Bragin, Anatol and Jand{\'o}, G{\'a}bor and N{\'a}dasdy, Zolt{\'a}n and Hetke, Jamille and Wise, K and Buzs{\'a}ki, Gy},
  journal={Journal of Neuroscience},
  volume={15},
  number={1},
  pages={47--60},
  year={1995},
  publisher={Soc Neuroscience}
}

@article{CsicsvariEtAl2003,
   author    = {Csicsvari, J. and Jamieson, B. and Wise, K. and Buzs\'aki, G.},
   title         = {Mechanisms of gamma oscillations in the hippocampus of the behaving rat}, 
   year       = {2003},
   journal   = {Neuron},
   volume  = {37},
   pages    = {311-322}
}

@article{ColginEtAl2009,
   author    = {Colgin, L. and Denninger, T. and Fyhn, M. and Hafting, T. and Bonnevie, T. and Jensen, O. and Moser, M. and Moser, E.},
   title         = {Frequency of gamma oscillations routes flow of information in the hippocampus}, 
   year       = {2009},
   journal   = {Nature},
   volume  = {462},
   pages    = {75-78}
}

@article{FriesEtAl2001,
   author    = {Fries, P. and Reynolds, J.H. and Rorie, A.E. and Desimone, R.},
   title         = {Modulation of oscillatory neuronal synchronization by selective visual attention}, 
   year       = {2001},
   journal   = {Science},
   volume  = {291},
   pages    = {1560-1563}
}

@article{WomelsdorfEtAl2007,
   author    = {Womelsdorf, T. and Schoffelen, J.M. and Oostenveld, R. and Singer, W. and Desimone, R. and Engel, A.K. and Fries, P.},
   title         = {Modulation of neuronal interactions through neuronal synchronization}, 
   year       = {2007},
   journal   = {Science},
   volume  = {316},
   pages    = {1609-1612}
}

@article{BroschEtAl2002,
  title={Stimulus-related gamma oscillations in primate auditory cortex},
  author={Brosch, Michael and Budinger, Eike and Scheich, Henning},
  journal={Journal of Neurophysiology},
  volume={87},
  number={6},
  pages={2715--2725},
  year={2002},
  publisher={American Physiological Society Bethesda, MD}
}

@article{bauer2006tactile,
  title={Tactile spatial attention enhances gamma-band activity in somatosensory cortex and reduces low-frequency activity in parieto-occipital areas},
  author={Bauer, Markus and Oostenveld, Robert and Peeters, Maarten and Fries, Pascal},
  journal={Journal of Neuroscience},
  volume={26},
  number={2},
  pages={490--501},
  year={2006},
  publisher={Soc Neuroscience}
}

@article{PesaranEtAl2002,
  title={Temporal structure in neuronal activity during working memory in macaque parietal cortex},
  author={Pesaran, Bijan and Pezaris, John S and Sahani, Maneesh and Mitra, Partha P and Andersen, Richard A},
  journal={Nature Neuroscience},
  volume={5},
  number={8},
  pages={805--811},
  year={2002},
  publisher={Nature Publishing Group}
}

@article{BuschmanMiller2007,
   author    = {Buschman, T.J. and Miller, E.K.},
   title         = {Top-down versus bottom-up control of attention in the prefrontal and posterior parietal cortices}, 
   year       = {2007},
   journal   = {Science},
   volume  = {315},
   pages    = {1860-1862}
}

@article{MedendorpEtAl2007,
  title={Oscillatory activity in human parietal and occipital cortex shows hemispheric lateralization and memory effects in a delayed double-step saccade task},
  author={Medendorp, W Pieter and Kramer, Geerten FI and Jensen, Ole and Oostenveld, Robert and Schoffelen, Jan-Mathijs and Fries, Pascal},
  journal={Cerebral Cortex},
  volume={17},
  number={10},
  pages={2364--2374},
  year={2007},
  publisher={Oxford University Press}
}

@article{GregorgiouEtAl2009,
   author    = {Gregoriou, G.G. and Gotts, S.J. and Zhou, H. and Desimone, R.},
   title         = {High-frequency, long-range coupling between prefrontal and visual cortex during attention}, 
   year       = {2009},
   journal   = {Science},
   volume  = {324},
   pages    = {1207-1210}
}

@article{SiegelEtAl2009,
  title={Phase-dependent neuronal coding of objects in short-term memory},
  author={Siegel, Markus and Warden, Melissa R and Miller, Earl K},
  journal={Proceedings of the National Academy of Sciences},
  volume={106},
  number={50},
  pages={21341--21346},
  year={2009},
  publisher={National Acad Sciences}
}

@article{SohalEtAl2009,
   author    = {Sohal, V.S. and Zhang, F. and Yizhar, O. and Deisseroth, K.},
   title         = {Parvalbumin neurons and gamma rhythms enhance cortical circuit performance}, 
   year       = {2009},
   journal   = {Nature},
   volume  = {459},
   pages    = {698-702}
}

@article{CanoltyEtAl2010,
  title={Oscillatory phase coupling coordinates anatomically dispersed functional cell assemblies},
  author={Canolty, Ryan T and Ganguly, Karunesh and Kennerley, Steven W and Cadieu, Charles F and Koepsell, Kilian and Wallis, Jonathan D and Carmena, Jose M},
  journal={Proceedings of the National Academy of Sciences},
  volume={107},
  number={40},
  pages={17356--17361},
  year={2010},
  publisher={National Acad Sciences}
}

@article{SigurdssonEtAl2010,
   author    = {Sigurdsson, T. and Stark, K.L. and Karayiorgou, M. and Gogos, J.A. and Gordon, J.A.},
   title         = {Impaired hippocampal-prefrontal synchrony in a genetic mouse model of schizophrenia}, 
   year       = {2010},
   journal   = {Nature},
   volume  = {464},
   pages    = {763-767}
}

@article{vanderMeerRedish2009,
  title={Low and high gamma oscillations in rat ventral striatum have distinct relationships to behavior, reward, and spiking activity on a learned spatial decision task},
  author={Van Der Meer, Matthijs AA and Redish, A David},
  journal={Frontiers in Integrative Neuroscience},
  volume={3},
  pages={9},
  year={2009},
  publisher={Frontiers}
}

@article{PopescuEtAl2009,
  title={Coherent gamma oscillations couple the amygdala and striatum during learning},
  author={Popescu, Andrei T and Popa, Daniela and Par{\'e}, Denis},
  journal={Nature Neuroscience},
  volume={12},
  number={6},
  pages={801--807},
  year={2009},
  publisher={Nature Publishing Group}
}

@article{azouz2000dynamic,
  title={Dynamic spike threshold reveals a mechanism for synaptic coincidence detection in cortical neurons in vivo},
  author={Azouz, Rony and Gray, Charles M},
  journal={Proceedings of the National Academy of Sciences},
  volume={97},
  number={14},
  pages={8110--8115},
  year={2000},
  publisher={National Acad Sciences}
}

@article{AzouzGray2003,
   author    = {Azouz, R. and Gray, C.M.},
   title         = {Adaptive coincidence detection and dynamic gain control in visual cortical neurons in vivo}, 
   year       = {2003},
   journal   = {Neuron},
   volume  = {37},
   pages    = {513-523}
}

@article{FrienEtAl2000,
  title={Fast oscillations display sharper orientation tuning than slower components of the same recordings in striate cortex of the awake monkey},
  author={Frien, Axel and Eckhorn, Reinhard and Bauer, Roman and Woelbern, Thomas and Gabriel, Andreas},
  journal={European Journal of Neuroscience},
  volume={12},
  number={4},
  pages={1453--1465},
  year={2000},
  publisher={Wiley Online Library}
}

@article{WomelsdorfEtAl2012,
  title={Orientation selectivity and noise correlation in awake monkey area V1 are modulated by the gamma cycle},
  author={Womelsdorf, Thilo and Lima, Bruss and Vinck, Martin and Oostenveld, Robert and Singer, Wolf and Neuenschwander, Sergio and Fries, Pascal},
  journal={Proceedings of the National Academy of Sciences},
  volume={109},
  number={11},
  pages={4302--4307},
  year={2012},
  publisher={National Acad Sciences}
}

@article{FriesEtAl2008,
  title={The effects of visual stimulation and selective visual attention on rhythmic neuronal synchronization in macaque area V4},
  author={Fries, Pascal and Womelsdorf, Thilo and Oostenveld, Robert and Desimone, Robert},
  journal={Journal of Neuroscience},
  volume={28},
  number={18},
  pages={4823--4835},
  year={2008},
  publisher={Soc Neuroscience}
}

@article{HenrieShapley2005,
  title={LFP power spectra in V1 cortex: the graded effect of stimulus contrast},
  author={Henrie, J Andrew and Shapley, Robert},
  journal={Journal of Neurophysiology},
  volume={94},
  number={1},
  pages={479--490},
  year={2005},
  publisher={American Physiological Society}
}

@article{LiuNewsome2006,
  title={Local field potential in cortical area MT: stimulus tuning and behavioral correlations},
  author={Liu, Jing and Newsome, William T},
  journal={Journal of Neuroscience},
  volume={26},
  number={30},
  pages={7779--7790},
  year={2006},
  publisher={Soc Neuroscience}
}

@Article{DiesmannEtAl1999,
   Author="Diesmann, M.  and Gewaltig, M. O.  and Aertsen, A. ",
   Title="{{S}table propagation of synchronous spiking in cortical neural networks}",
   Journal="Nature",
   Year="1999",
   Volume="402",
   Pages="529--533",
}

@Article{RayMaunsell2015,
  title={Do gamma oscillations play a role in cerebral cortex?},
  author={Ray, Supratim and Maunsell, John HR},
  journal={Trends in Cognitive Sciences},
  volume={19},
  number={2},
  pages={78--85},
  year={2015},
  publisher={Elsevier}
}

@Article{KeeleyEtAl2019,
  title={Firing rate models for gamma oscillations},
  author={Keeley, Stephen and Byrne, {\'A}ine and Fenton, Andr{\'e} and Rinzel, John},
  journal={Journal of Neurophysiology},
  volume={121},
  number={6},
  pages={2181--2190},
  year={2019},
  publisher={American Physiological Society Bethesda, MD}
}

@Article{Fries2009,
  title={Neuronal gamma-band synchronization as a fundamental process in cortical computation},
  author={Fries, Pascal},
  journal={Annual Review of Neuroscience},
  volume={32},
  pages={209--224},
  year={2009},
  publisher={Annual Reviews}
}

@article{borgers2003synchronization,
  title={Synchronization in networks of excitatory and inhibitory neurons with sparse, random connectivity},
  author={B{\"o}rgers, Christoph and Kopell, Nancy},
  journal={Neural Computation},
  volume={15},
  number={3},
  pages={509--538},
  year={2003},
  publisher={MIT Press}
}

@article{ashwin2016mathematical,
  title={Mathematical frameworks for oscillatory network dynamics in neuroscience},
  author={Ashwin, Peter and Coombes, Stephen and Nicks, Rachel},
  journal={The Journal of Mathematical Neuroscience},
  volume={6},
  number={1},
  pages={2},
  year={2016},
  publisher={Springer}
}

@article{bick2020understanding,
  title={Understanding the dynamics of biological and neural oscillator networks through exact mean-field reductions: a review},
  author={Bick, Christian and Goodfellow, Marc and Laing, Carlo R and Martens, Erik A},
  journal={The Journal of Mathematical Neuroscience},
  volume={10},
  number={1},
  pages={1--43},
  year={2020},
  publisher={SpringerOpen}
}

@article{Basar2013,
title = "A review of gamma oscillations in healthy subjects and in cognitive impairment",
journal = "International Journal of Psychophysiology",
volume = "90",
number = "2",
pages = "99 - 117",
year = "2013",
issn = "0167-8760",
doi = "https://doi.org/10.1016/j.ijpsycho.2013.07.005",
url = "http://www.sciencedirect.com/science/article/pii/S0167876013002134",
author = "Erol Başar"
}

@Article{TallonBaudry2009,
   author={Tallon-Baudry, Catherine},
   Title="{{T}he roles of gamma-band oscillatory synchrony in human visual cognition}",
   Journal="Frontiers in Bioscience (Landmark Ed)",
   Year="2009",
   Volume="14",
   Pages="321--332",
   Month="1"
}

@Article{LogothetisEtAl2001,
   Author="Logothetis, N. K.  and Pauls, J.  and Augath, M.  and Trinath, T.  and Oeltermann, A. ",
   Title="{{N}europhysiological investigation of the basis of the f{M}{R}{I} signal}",
   Journal="Nature",
   Year="2001",
   Volume="412",
   Number="6843",
   Pages="150--157",
   Month="7"
}

@Article{TraubEtAl2005,
  title={Single-column thalamocortical network model exhibiting gamma oscillations, sleep spindles, and epileptogenic bursts},
  author={Traub, Roger D and Contreras, Diego and Cunningham, Mark O and Murray, Hilary and LeBeau, Fiona EN and Roopun, Anita and Bibbig, Andrea and Wilent, W Bryan and Higley, Michael J and Whittington, Miles A},
  journal={Journal of Neurophysiology},
  volume={93},
  number={4},
  pages={2194--2232},
  year={2005},
  publisher={American Physiological Society}
}

@Article{SongEtAl2005,
  title={Highly nonrandom features of synaptic connectivity in local cortical circuits},
  author={Song, Sen and Sj{\"o}str{\"o}m, Per Jesper and Reigl, Markus and Nelson, Sacha and Chklovskii, Dmitri B},
  journal={PLoS Biology},
  volume={3},
  number={3},
  pages={e68},
  year={2005},
  publisher={Public Library of Science}
}

@Article{Wang2010,
  title={Neurophysiological and computational principles of cortical rhythms in cognition},
  author={Wang, Xiao-Jing},
  journal={Physiological Reviews},
  volume={90},
  number={3},
  pages={1195--1268},
  year={2010},
  publisher={American Physiological Society Bethesda, MD}
}

@Article{BuzsakiWang2012,
  title={Mechanisms of gamma oscillations},
  author={Buzs{\'a}ki, Gy{\"o}rgy and Wang, Xiao-Jing},
  journal={Annual Review of Neuroscience},
  volume={35},
  pages={203--225},
  year={2012},
  publisher={Annual Reviews}
}

@Article{SiegelDonnerEngel2012,
  title={Spectral fingerprints of large-scale neuronal interactions},
  author={Siegel, Markus and Donner, Tobias H and Engel, Andreas K},
  journal={Nature Reviews Neuroscience},
  volume={13},
  number={2},
  pages={121--134},
  year={2012},
  publisher={Nature Publishing Group}
}

@Article{ZhangRangan2015,
   Author="Zhang, J. W.  and Rangan, A. V. ",
   Title="{{A} reduction for spiking integrate-and-fire network dynamics ranging from homogeneity to synchrony}",
   Journal="Journal of Computational Neuroscience",
   Year="2015",
   Volume="38",
   Number="2",
   Pages="355--404",
   Month="04"
}

@Article{ZhangZhouEtAl2014,
   Author="Zhang, J.  and Zhou, D.  and Cai, D.  and Rangan, A. V. ",
   Title="{{A} coarse-grained framework for spiking neuronal networks: between homogeneity and synchrony}",
   Journal="Journal of Computational Neuroscience",
   Year="2014",
   Volume="37",
   Number="1",
   Pages="81--104",
   Month="08"
}

@Article{ZhangNewhallEtAl2014,
   Author="Zhang, J.  and Newhall, K.  and Zhou, D.  and Rangan, A. ",
   Title="{{D}istribution of correlated spiking events in a population-based approach for {I}ntegrate-and-{F}ire networks}",
   Journal="Journal of Computational Neuroscience",
   Year="2014",
   Volume="36",
   Number="2",
   Pages="279--295",
   Month="04"
}

@Article{RanganYoung2013b,
   Author="Rangan, A. V.  and Young, L. S. ",
   Title="{{E}mergent dynamics in a model of visual cortex}",
   Journal="Journal of Computational Neuroscience",
   Year="2013",
   Volume="35",
   Number="2",
   Pages="155--167",
   Month="10"
}

@Article{RanganYoung2013a,
   Author="Rangan, A. V.  and Young, L. S. ",
   Title="{{D}ynamics of spiking neurons: between homogeneity and synchrony}",
   Journal="Journal of Computational Neuroscience",
   Year="2013",
   Volume="34",
   Number="3",
   Pages="433--460",
   Month="06"
}

@Article{Laing2018,
   Author="Laing, C. R. ",
   Title="{{T}he dynamics of networks of identical theta neurons}",
   Journal="Journal of Mathematical Neuroscience",
   Year="2018",
   Volume="8",
   Number="1",
   Pages="4",
   Month="02"
}

% \appendix
\newpage
\section*{Supplementary Materials}
\renewcommand{\thesubsection}{S\arabic{subsection}} 
\renewcommand{\thefigure}{S\arabic{figure}} 
\setcounter{figure}{0}

\subsection{Low-Dimension Manifolds}
\begin{figure}[h]
  \begin{center}
    %% \captionsetup{type=figure} 
    \includegraphics[width=\textwidth]{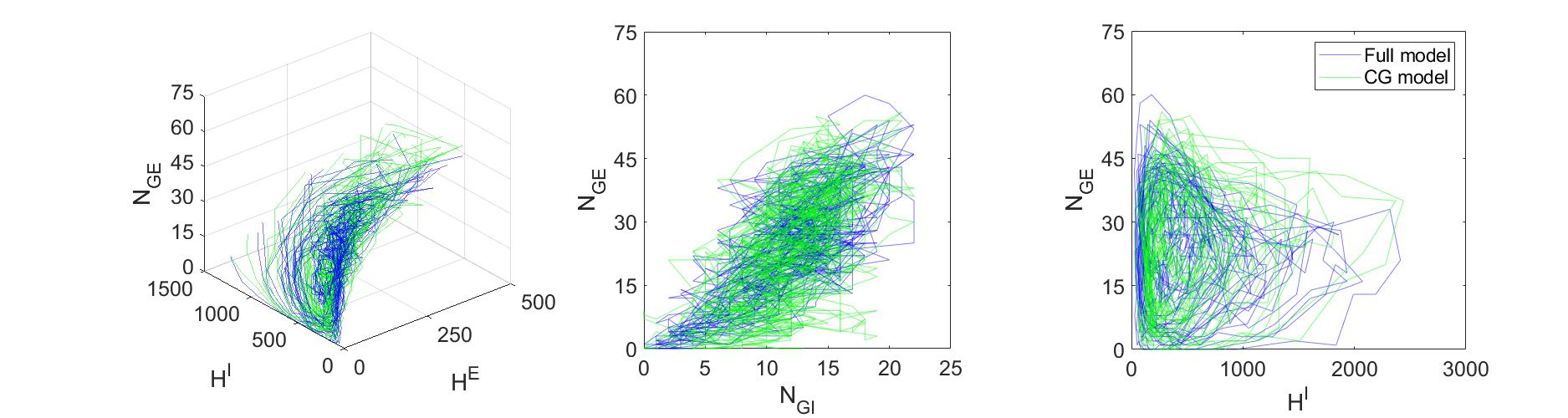}
    \caption{MIF and CG models produce trajectories on similar surfaces.  Left: The projection on subspace $(N_{GE},H^{E},H^{I})$; Middle: The projection on subspace $(N_{GE},N_{GI})$; Right: The projection on subspace $(N_{GE},H^{I})$. } 
    \label{FigS1: MIF_CG}
  \end{center}
\end{figure}

\heading{Trajectories of MIF and CG Models.} Although MIF and CG models exhibit subtle difference in the statistics of the dynamics, we find that their trajectories in state space $(N_{GE},N_{GI},H^E,H^I)$ stay close to similar manifolds (Fig.~\ref{FigS1: MIF_CG}). 

%%%%%%%%%%%%%%%%%%%%%%%%%%%%%%%%%%%%%%%%%%%%%%%%%%%%%%%%%%% FigS1
%%%%%%%%%%%%%%%%%%%%%%%%%%%%%%%%%%%%%%%%%%%%%%%%%%%%%%%%%%% FigS1

\begin{figure}[h]
  \begin{center}
    %% \captionsetup{type=figure} 
    \includegraphics[width=\textwidth]{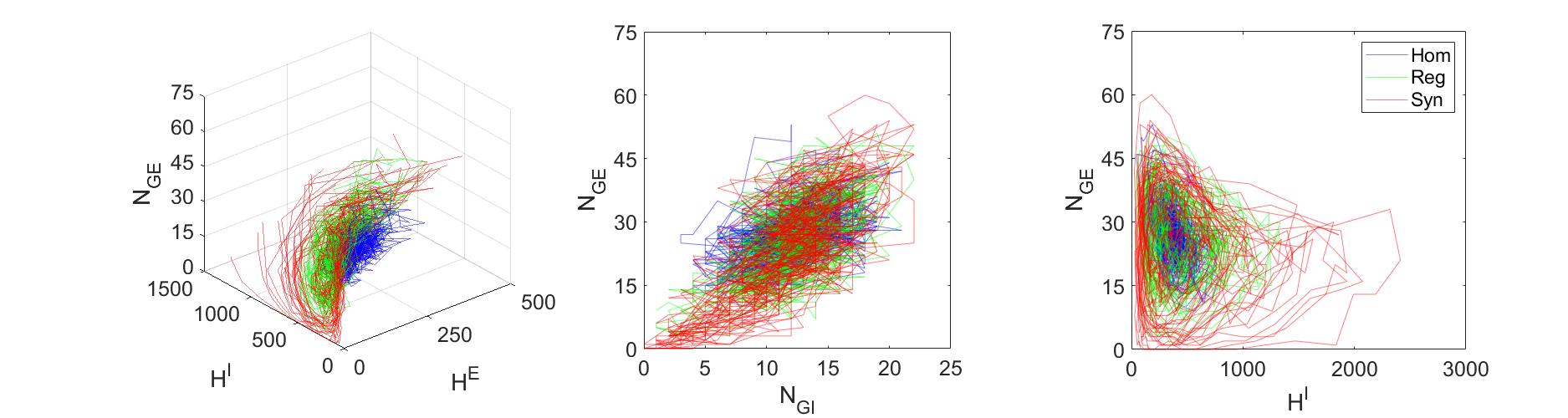}
    \caption{Full model trajectories of three regimes (Syn, Reg, and Hom).  Left: The projection on subspace $(N_{GE},H^{E},H^{I})$; Middle: The projection on subspace $(N_{GE},N_{GI})$; Right: The projection on subspace $(N_{GE},H^{I})$.} 
    \label{FigS2: Three_regimes}
  \end{center}
\end{figure}

\heading{MIF Trajectories in Syn, Reg, and Hom Regimes.} The low-dimensional manifolds in state spaces are powerful tools to visualize and analyze gamma dynamics. Here, we are interested in the comparison of manifolds for these three regimes studied in the manuscript. In the Syn regime, $X$ has a faster speed to consume the pending $E$-kicks (compared to Hom and Reg), i.e., $E$-neurons are faster recruited by the recurrent excitation, resulting in larger MFEs. These features are reflected by the full-model trajectories in the three different regimes (Fig.~\ref{FigS2: Three_regimes}).

%%%%%%%%%%%%%%%%%%%%%%%%%%%%%%%%%%%%%%%%%%%%%%%%%%%%%%%%%%% FigS2
%%%%%%%%%%%%%%%%%%%%%%%%%%%%%%%%%%%%%%%%%%%%%%%%%%%%%%%%%%% FigS2

\newpage
\subsection{A Larger MIF Network}
\begin{figure}[h]
  \begin{center}
    %% \captionsetup{type=figure} 
    \includegraphics[width=\textwidth]{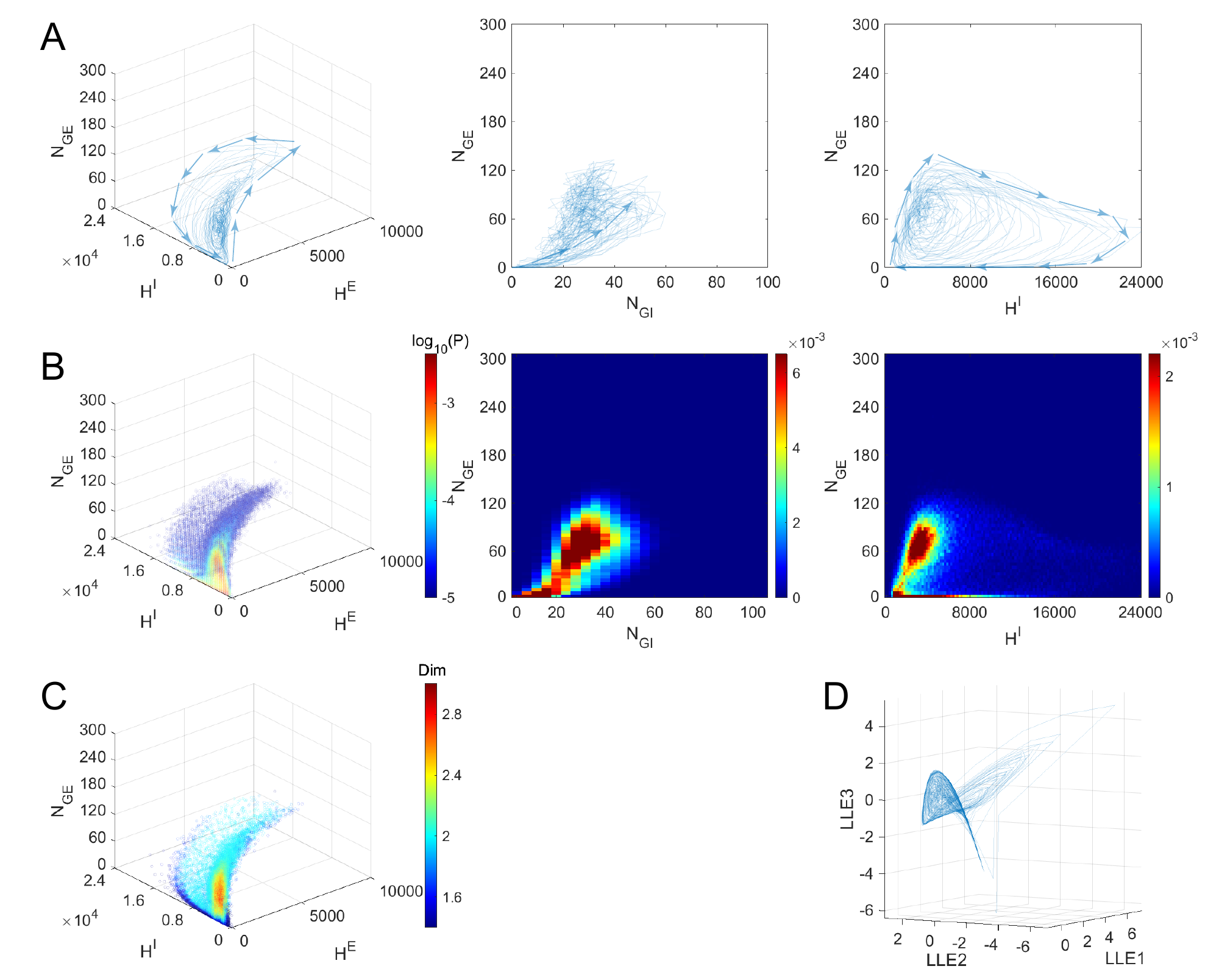}
    \caption{Gamma dynamics of a 400-neuron MIF network restricted on a low-dimensional manifold. \textbf{A-D} are similar to \textbf{A, B, E, F} of Fig.~\ref{Fig6: Manifold}.} 
    \label{FigS3: Larger_Network}
  \end{center}
\end{figure}

We repeat the model reduction for a 400-neuron MIF network, and collect the state space variables $(N_{GE},N_{GI},H^{E},H^{I})$ from a 100-second simulation (Fig.~\ref{FigS3: Larger_Network}).  A similar manifold is revealed by the trajectories in the state space, and also verified by the mass of trajectory density (Fig.~\ref{FigS3: Larger_Network}AB). The two-dimensional local structure is also verified by the local dimensionality estimation and local linear embedding (Fig.~\ref{FigS3: Larger_Network}CD). For this network, we simulate the Syn regime with parameters identical to \cite{li2019well}.  

%%%%%%%%%%%%%%%%%%%%%%%%%%%%%%%%%%%%%%%%%%%%%%%%%%%%%%%%%%% FigS3
%%%%%%%%%%%%%%%%%%%%%%%%%%%%%%%%%%%%%%%%%%%%%%%%%%%%%%%%%%% FigS3

\newpage
\subsection{Issues of Model Reduction}
\heading{Selection of the cutoff.} How faithful the RN models capture the full-model dynamics depends on the selection of cutoff $V^c$.  Our assumption of RN is that the gamma dynamics is mostly sensitive to the subthreshold distribution of neurons about to fire. On the other hand, a high $V^c$ provides an accurate description of gate neurons, but very roughly for the base neurons, which is a potential issue for the RN model. For the RN model in this paper, we tuned the selection of the cutoff systematically and finally choose $V^c = 40$. Here we present the firing rates and raster plots of the RN model when $V^c = 70$ and $40$, where the former results in much larger firing rates and MFEs. 

%\heading{Ref}

%%%%%%%%%%%%%%%%%%%%%%%%%%%%%%%%%%%%%%%%%%%%%%%%%%%%%%%%%%% FigS4
\begin{figure}[h]
  \begin{center}
    %% \captionsetup{type=figure} 
    \includegraphics[width=\textwidth]{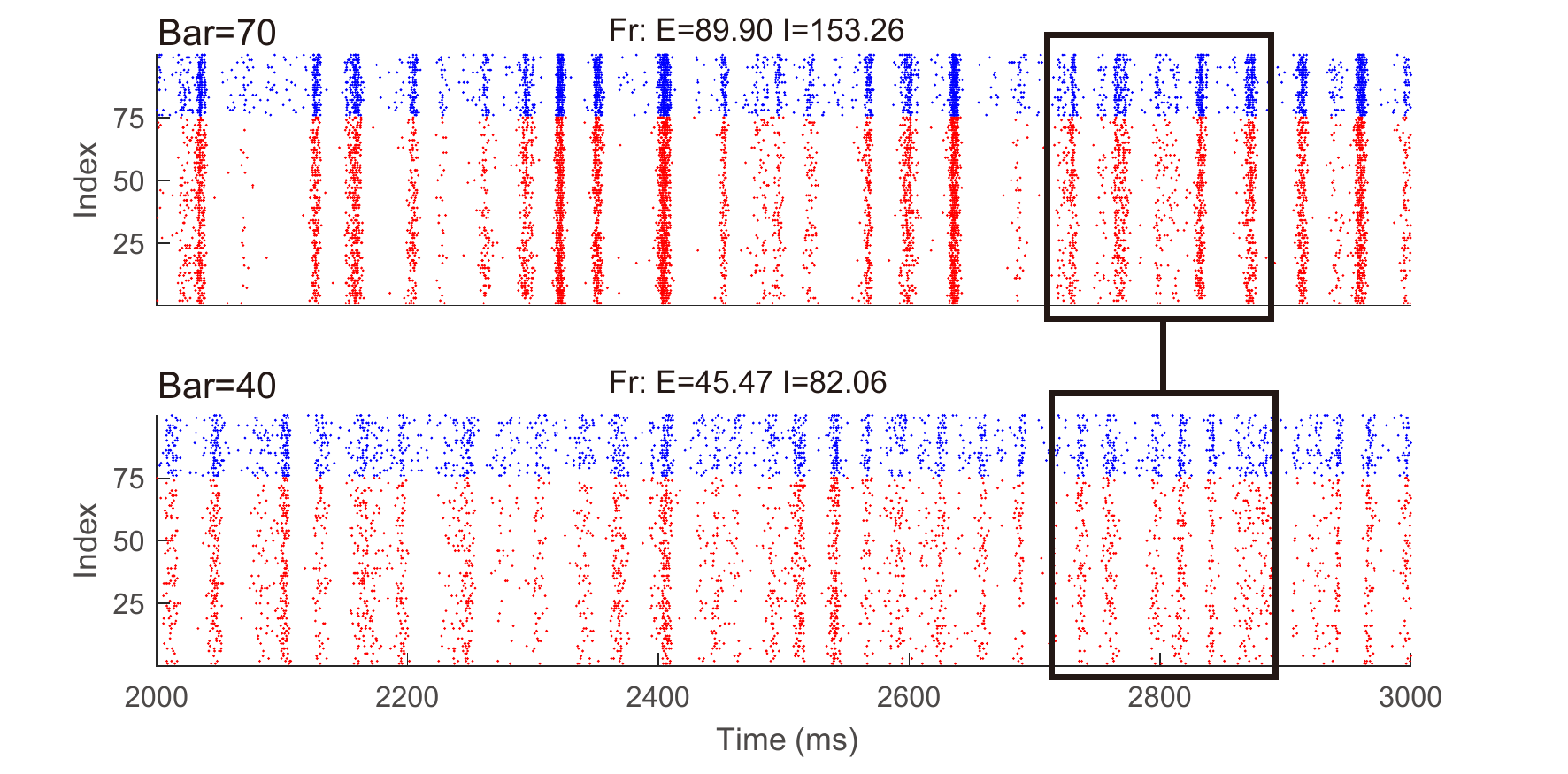}
    \caption{Raster plots of different cutoffs $V^c$ in the RN models. Upper: $V^c =  70$; Lower: $V^c = 40$.} 
    \label{FigS4: Reduction Issues}
  \end{center}
\end{figure}
%%%%%%%%%%%%%%%%%%%%%%%%%%%%%%%%%%%%%%%%%%%%%%%%%%%%%%%%%%% FigS4

\heading{The duration of MFEs.} The MFEs of the reduced models have longer duration than the full model, possibly due to the slower probability flows since we combine multiple states from the full model as one in the reduced ones. 

%%%%%%%%%%%%%%%%%%%%%%%%%%%%%%%%%%%%%%%%%%%%%%%%%%%%%%%%%%% FigS5
\begin{figure}[h]
  \begin{center}
    %% \captionsetup{type=figure} 
    \includegraphics[width=0.7\textwidth]{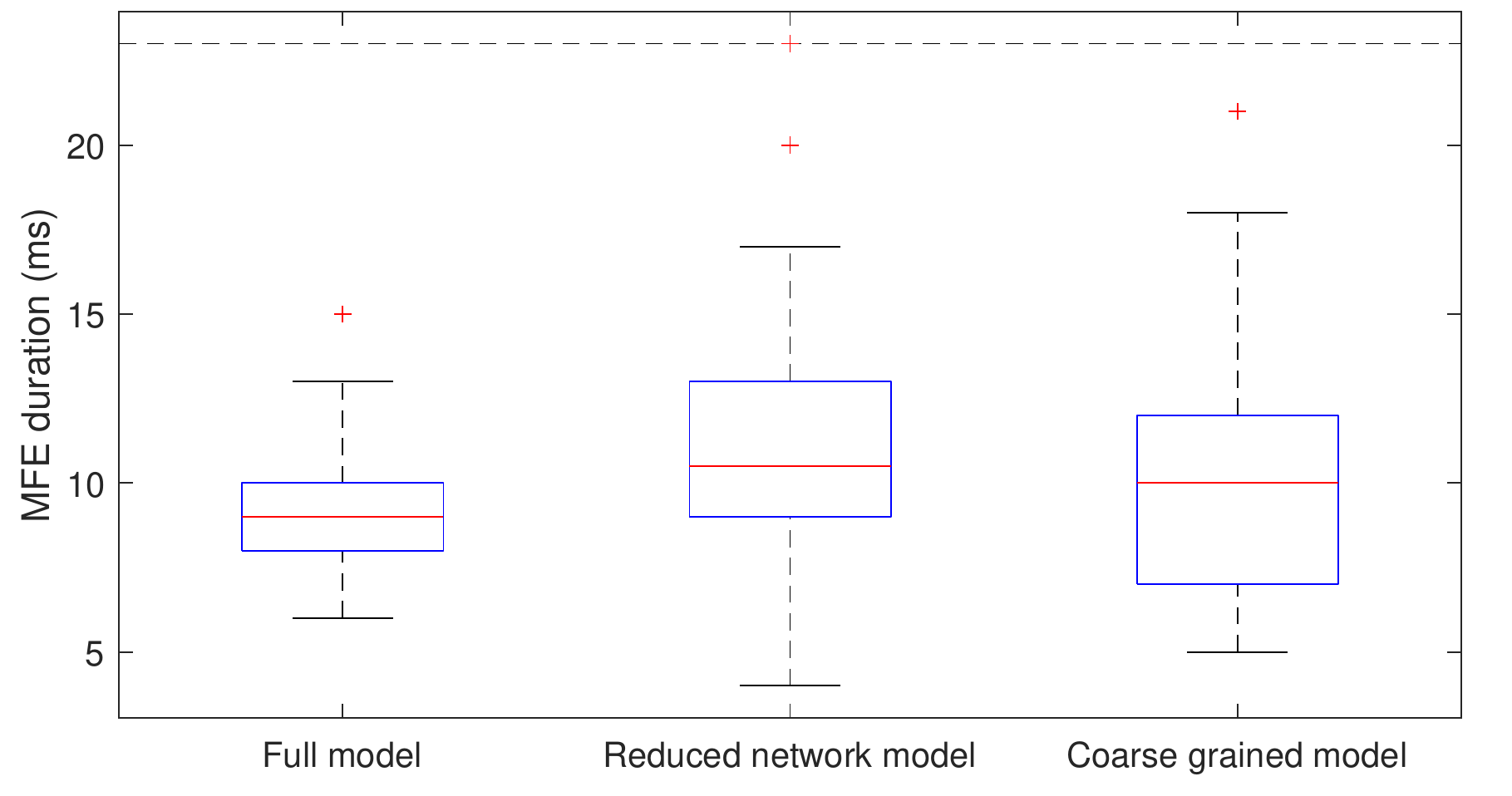}
    \caption{The duration of MFEs in different models. For each model, the mean (red line), first and third quartile (lower and upper bounds of the blue box), minimum and maximum (black lines), and outliers (red cross) are indicated in the box plot. The duration of MFEs for the RN and CG models are significantly higher than the full model (p = 0.004 and 0.038).} 
    \label{FigS5: MFE duration}
  \end{center}
\end{figure}
%%%%%%%%%%%%%%%%%%%%%%%%%%%%%%%%%%%%%%%%%%%%%%%%%%%%%%%%%%% FigS5

\heading{A 5-variable CG model.} The 4-variable CG model combines $E$-kick pools of all neurons together, which may cause a problem since the pending $E$-kicks for $E$ neurons and $I$ neurons are consumed by different speeds.  Hence we come up with a 5-variable version were $H_{EE}$ and $H_{IE}$ are counted separately.

%%%%%%%%%%%%%%%%%%%%%%%%%%%%%%%%%%%%%%%%%%%%%%%%%%%%%%%%%%% FigS6
\begin{figure}[h]
  \begin{center}
    %% \captionsetup{type=figure} 
    \includegraphics[width=0.5\textwidth]{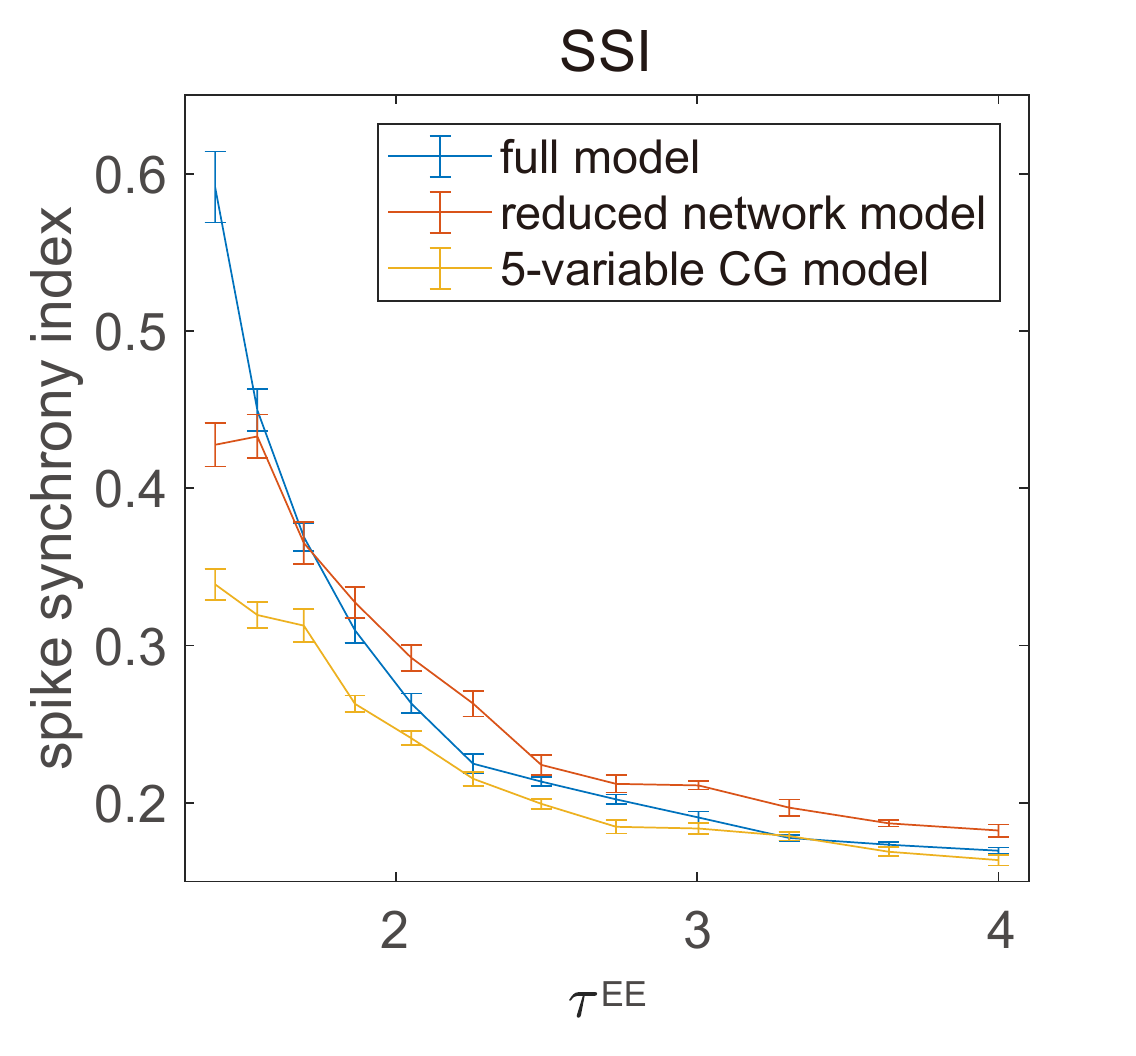}
    \caption{Degree of synchrony increases when $\tau^{EE}$ decreases. Reprinted from Fig.~\ref{Fig5: Gamma Features}B, but the 4-variable CG model is replaced by a 5-variable version.} 
    \label{FigS6: 5-Variable CG}
  \end{center}
\end{figure}
%%%%%%%%%%%%%%%%%%%%%%%%%%%%%%%%%%%%%%%%%%%%%%%%%%%%%%%%%%% FigS6

\subsection{Incorporate More Complicate Setup: NMDA Receptor and others}
The model reduction introduced in this paper can be easily generalized for more complicated network setup. We present one example here.

In the main text, we only considered two synapse types (or more frankly, two synaptic time scales): AMPA and GABA, in this paper. To incorporate more types of synapse, such as NMDA (~80 ms), we can introduce one more pending-kick pools $H^{\mbox{nmda}}_i$ for neuron $i$ in the MIF and RN models, and an additional pending-kick pool $H^{\mbox{nmda}}$ for the whole network. We note that this idea can be extended to multiple, heterogeneously coupled populations. In general, we may need to use states beyond base and gate, and the relevant pending-kick pools, as synaptic couplings become heterogeneous.

%\subsection*{Reduction for Leaky-Integrate-Fire Models}

\end{document}